\documentclass[american,aps,pre,reprint,tightenlines,superscriptaddress,twocolumn,twoside,longbibliography]{revtex4-2}

\pdfoutput=1

\usepackage[utf8]{inputenc}

\usepackage[unicode=true,pdfusetitle, bookmarks=true,bookmarksnumbered=false,bookmarksopen=false, breaklinks=false,pdfborder={0 0 0},backref=false,colorlinks=false] {hyperref}
\hypersetup{ colorlinks,linkcolor=myurlcolor,citecolor=myurlcolor,urlcolor=myurlcolor}
\usepackage{braket,colortbl,amsthm,amsmath,amssymb,dsfont}
\definecolor{myurlcolor}{rgb}{0,0,0.7}
\usepackage{graphics,graphicx,relsize}
\usepackage{color}
\usepackage{XCharter}
\usepackage[T1]{fontenc}

\usepackage{enumitem}  

\setlist[itemize]{left=0pt, labelindent=0pt}

\usepackage{graphicx}
\usepackage{subcaption}
\usepackage{url}
\usepackage[font=small,labelfont=bf,justification=raggedright]{caption}

\usepackage{mathptmx}
\usepackage[left=1.6cm,right=1.6cm,top=2.45cm,bottom=2.45cm]{geometry}
\usepackage{bigints}
\usepackage{accents}

\usepackage{multirow}
\usepackage{multirow}
\usepackage[table,xcdraw]{xcolor}

\usepackage{float} 

\theoremstyle{plain}

\def\bea{\begin{eqnarray}}
\def\eea{\end{eqnarray}}
\def\ba{\begin{array}}
\def\ea{\end{array}}

\def\beq{\begin{equation}}
\def\eeq{\end{equation}}

\usepackage[normalem]{ulem}
 
\makeatletter

\begin{document}
\begingroup
\maketitle
\endgroup

\title{High-Efficiency Three-Stroke Quantum Isochoric Heat Engine: From Infinite Potential Wells to Magic-Angle Twisted Bilayer Graphene}

\author{Hadi Mohammed Soufy}
\email{hm.soufy@niser.ac.in}
\affiliation{School of Physical Sciences, National Institute of Science Education and Research, HBNI, Jatni-752050, India}
\affiliation{Homi Bhabha National Institute, Training School Complex, Anushakti Nagar, Mumbai, 400094, India}

\author{Colin Benjamin}
\email{colin.nano@gmail.com}
\affiliation{School of Physical Sciences, National Institute of Science Education and Research, HBNI, Jatni-752050, India}
\affiliation{Homi Bhabha National Institute, Training School Complex, Anushakti Nagar, Mumbai, 400094, India}

\begin{abstract}
We introduce a three-stroke quantum isochoric cycle that functions as a heat engine operating between two thermal reservoirs. Implemented for a particle confined in a one-dimensional infinite potential well, the cycle’s performance is benchmarked against the classical three-stroke triangular and isochoric engines. We find that the quantum isochoric cycle achieves a higher efficiency than both classical counterparts and also surpasses the efficiency of the recently proposed three-stroke quantum isoenergetic cycle. Owing to its reduced number of strokes, the design substantially lowers control complexity in nanoscale thermodynamic devices, offering a more feasible route to experimental realization compared to conventional four-stroke architectures. We further evaluate the cycle in graphene-based systems under an external magnetic field, including monolayer graphene (MLG), AB-stacked bilayer graphene (BLG), and twisted bilayer graphene (TBG) at both magic and non-magic twist angles. Among these platforms, magic-angle twisted bilayer graphene (MATBG) attains the highest efficiency at fixed work output, highlighting its promise for quantum thermodynamic applications.
\end{abstract}
\maketitle
\clearpage  

\section{Introduction}

The intersection of quantum mechanics and thermodynamics has led to a rapidly developing field focused on microscopic quantum thermodynamic devices (QTDs) \cite{uzdin2015equivalence,kosloff2013quantum,bhattacharjee2021quantum}. Operating at the scale of individual atoms and engineered quantum systems, these devices challenge classical intuition by revealing how coherence, discreteness, and quantum statistics influence energy-conversion processes \cite{vinjanampathy2016quantum,millen2016perspective,brask2015autonomous,myers2022quantum}. Over the past decade, QTDs have been realized experimentally in a wide range of platforms, including trapped ions \cite{peterson2019experimental,rossnagel2016single}, superconducting circuits \cite{uusnakki2025superconducting}, nitrogen-vacancy centers in diamond \cite{klatzow2019quantum}, ultracold atomic ensembles \cite{bouton2021quantum}, and optomechanical systems \cite{zhang2020optomechanical}. These breakthroughs demonstrate that thermodynamic cycles, traditionally associated with macroscopic engines, can be meaningfully implemented at the quantum scale, often with enhanced control and access to operational regimes without classical analogues. Designing such machines requires balancing heat exchange and work extraction across discrete thermodynamic strokes \cite{quan2007quantum,quan2009quantum}, with the number and nature of the strokes dictating how energy flow is orchestrated. Understanding how each stroke shapes performance is therefore essential for harnessing the capabilities of quantum materials. Although four-stroke cycles form the conventional framework for analyzing work and efficiency, many experimental platforms struggle to realize all four operations with high precision due to decoherence, noise, and limited control resources \cite{deffner2019quantum,abah2014efficiency}. Reducing the cycle to three strokes provides a practical alternative that retains essential thermodynamic features while significantly simplifying implementation, making three-stroke models both conceptually insightful and experimentally relevant \cite{koyanagi2022laws,lisboa2022experimental}.

In this work, we begin by analyzing two classical three-stroke engines, the triangular cycle and the isochoric cycle, to establish a baseline for comparison. We then introduce a quantum analogue by considering a particle confined in a one-dimensional infinite potential well (IPW) and implementing two distinct quantum three-stroke cycles: a quantum isochoric cycle operating between two thermal baths, and a quantum isoenergetic cycle, originally proposed in Ref.~\cite{ou2016exotic}, which operates by maintaining constant internal energy while interacting with a single bath. By systematically comparing their thermodynamic behavior, we show that the quantum isochoric cycle attains both higher efficiency and greater work output than the quantum isoenergetic cycle under identical operating conditions. Building on this insight, we extend our analysis to graphene-based platforms: monolayer graphene (MLG), AB Bernal-stacked bilayer graphene (BLG), and twisted bilayer graphene (TBG) at both magic and non-magic twist angles, in the presence of a perpendicular magnetic field. Implementing the proposed quantum isochoric cycle in these systems allows us to evaluate and compare their performance as quantum heat engines.

Section~\ref{II} summarizes the thermodynamic foundations relevant to our analysis, including the conditions for feasibility of a cycle and the operational regimes of quantum heat engines. The specific cycles investigated in this work are then introduced in the subsequent subsections: Section~\ref{IIA} reviews the classical triangular cycle; Section~\ref{IIB} presents the classical isochoric cycle; Section~\ref{IIC} details the proposed three-stroke quantum isochoric cycle; and Section~\ref{IID} discusses the three-stroke quantum isoenergetic cycle. For each case, we describe the implementation for a particle in an IPW, analyze compliance with the thermodynamic laws, and compare their performance metrics. Section~\ref{III} provides the theoretical background for graphene-based systems—MLG, BLG, and TBG at various twist angles—under an external magnetic field, while Section~\ref{IIIA} applies the quantum isochoric cycle to these platforms. Our main results and comparative analysis are presented in Section~\ref{IV}, followed by concluding remarks and potential experimental realizations in Section~\ref{V}.

\section{Thermodynamic Cycle}
\label{II}

To understand the performance of a system as a thermodynamic device, we analyze its behavior over a complete thermodynamic cycle. A cycle consists of a series of distinct thermodynamic strokes, that finally return the system to its initial state. The net work output and efficiency, which quantify the device's utility, are derived from the heat exchanged with the thermal reservoirs during these strokes \cite{vinjanampathy2016quantum,quan2007quantum,quan2009quantum,landau2013statistical}. For a thermodynamic stroke, the first law of thermodynamics states,
\begin{equation}
    \Delta U = Q - W,
    \label{eq:firstlaw}
\end{equation}
where \(Q\) is the heat supplied to the system, \(W\) is the work done by the system and \(\Delta U\) is the internal energy change. At the end of a cycle, \(\Delta U = 0\); hence,
\begin{equation}
    W_1 + W_2 + \dots = W = Q_1 + Q_2 + \dots
    \label{eq:cycle formula}
\end{equation}
\(W_1, W_2, \dots\) and \(Q_1, Q_2, \dots\) are the work and heat exchanged in the individual thermodynamic strokes. For a system undergoing a thermodynamic cycle across two reservoirs at temperatures \(T_h\) and \(T_c\), the heat exchanged with them is denoted by \(Q_{\text{hot}}\) and \(Q_{\text{cold}}\), respectively. Eq.~\eqref{eq:cycle formula} then reduces to,
\begin{equation}
    W = Q_{\text{hot}} + Q_{\text{cold}}.
    \label{eq:work-heat}
\end{equation}

The second law of thermodynamics requires total entropy change \(\Delta S_\text{tot}\) over a cycle be non-negative, which is a necessary condition for physical realizability \cite{landi2021irreversible,vinjanampathy2016quantum,landau2013statistical}. For a system undergoing a thermodynamic stroke while coupled to baths, this is expressed as,
\begin{equation}
    \Delta S_\text{tot} = \Delta S_{\text{sys}}+\Delta S_{\text{bath}} \ge 0,
\end{equation}  

Here, $\Delta S_{\text{sys}}$ denotes the entropy change of the system, and $\Delta S_{\text{bath}}$ corresponds to that of the baths during the stroke. For a reversible process $\Delta S_{\text{tot}}=0$. For a system undergoing a thermodynamic cycle, which may be reversible or irreversible, between two heat baths at temperatures \(T_h\) and \(T_c\) (\(T_h > T_c\)) \cite{vinjanampathy2016quantum,landi2021irreversible}, we have,
\vspace{-2em}

\begin{equation}
   \Delta S_\text{sys}=0,\quad \text{and}\quad\Delta S_{\text{bath}} = -\frac{Q_\text{hot}}{T_h}-\frac{Q_\text{cold}}{T_c},\nonumber
\end{equation}  
\vspace{-2em}
\begin{flalign}
&\text{Thus,} \quad \quad\quad\Delta S_{\text{tot}} = -\frac{Q_\text{hot}}{T_h}-\frac{Q_\text{cold}}{T_c} \ge 0,&
\label{eq:secondall}
\end{flalign}
\vspace{-2em}
\begin{flalign}
&\text{which leads to the condition:}\quad\quad
    \frac{Q_\text{hot}}{T_h}+\frac{Q_\text{cold}}{T_c} \le 0.
    \label{eq:secondlawcond}&
\end{flalign}

 A heat engine is characterized by \(W > 0\), \(Q_{\text{hot}} > 0\), and \(Q_{\text{cold}} < 0\), which corresponds to heat \(Q_{\text{hot}}\) being absorbed from the hot bath and \(Q_{\text{cold}}\) being rejected to the cold bath while performing work \(W\) \cite{soufy2025flatband,scharf2020topological}.






\begin{figure}[]
    \centering
    \includegraphics[width=1\linewidth]{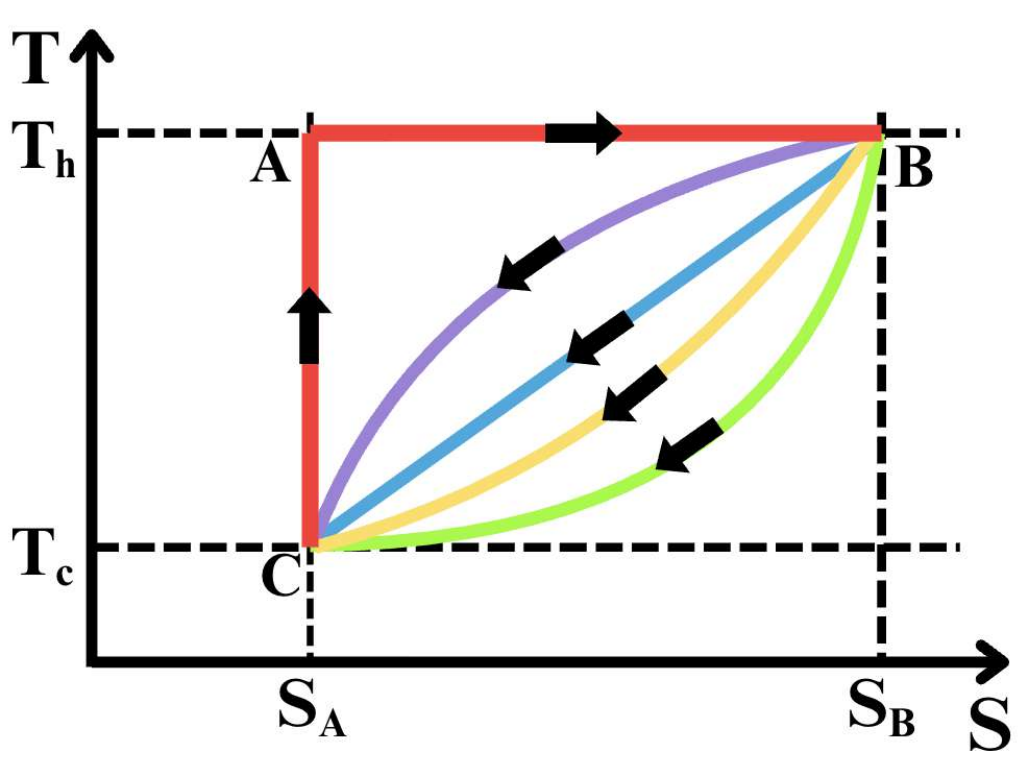}
    \caption{Entropy - Temperature (T-S) diagrams for three thermodynamic cycles. Shown are: the classical triangular cycle \(A \to B \xrightarrow{\text{Blue}} C \to A\) \cite{rau2017statistical}, the classical isochoric cycle \(A \to B \xrightarrow{\text{Yellow}} C \to A\), the quantum isochoric cycle \(A \to B \xrightarrow{\text{Green}} C \to A\), and the quantum isoenergetic cycle \(A \to B \xrightarrow{\text{Purple}} C \to A\) \cite{ou2016exotic}. The classical triangular, classical isochoric and quantum isochoric cycles operate between two baths at fixed temperatures \(T_h\) and \(T_c\), while the quantum isoenergetic cycle operates with a single bath at \(T_h\), where the effective cold temperature \(T_c\) is not fixed but is determined by other parameters.}
    \label{Fig1}
\end{figure}

\subsection{Three-Stroke Classical Triangular Cycle}
\label{IIA}



In Fig.~\ref{Fig1}, \(A\xrightarrow{}B\xrightarrow{\text{Blue}}C\xrightarrow{}A\) illustrates three-stroke triangular cycle. It consists of an isothermal stroke (\(A\xrightarrow{}B\)), an adiabatic stroke (\(C\xrightarrow{}A\)), and an irreversible stroke that connects the endpoints of the first two (\(B \xrightarrow{\text{Blue}}C\)) \cite{rau2017statistical,landau2013statistical}. For such a cycle, the performance coefficients depend solely on the temperature gradient between the thermal baths and is irreversible \cite{rau2017statistical,landau2013statistical}(see, Eq.\eqref{eq:12}).

The system exchanges heat \(Q_{\text{hot}}\) with the hot bath during the isothermal stroke and heat \(Q_{\text{cold}}\) at the end of the irreversible stroke. The system is then brought back to its initial state by an adiabatic process. \(Q_{\text{hot}}\) corresponds to the area under line AB, \(Q_{\text{cold}}\) is the negative of the area under blue line BC. Thus,

\begin{equation}
\begin{split}
     Q_{\text{hot}} = T_h(S_B - S_A),\quad \quad Q_{\text{cold}} = \frac{1}{2}(T_h + T_c)(S_A - S_B),
     \label{heatclassical}
\end{split}
\end{equation}

\noindent and work done \(W_\text{CT}\) is the area of the cycle ABC (with the blue BC line), 
\begin{equation}
    W_\text{CT}= \frac{1}{2} \left(T_h-T_c\right)\left(S_B-S_A\right) = \frac{W_{\text{c}}}{2},
    \label{eq:Workmonotri}
\end{equation}

\noindent where, \(W_c\) is the work done during a Carnot cycle operating between temperatures \(T_h\) and \(T_c\). For \(N\) atoms of a classical ideal gas, work done during a Carnot cycle ($W_c$), is given by \cite{callen2006thermodynamics,landau2013statistical},
\vspace{-2em}
\begin{equation}
    W_c=\frac{Nk_B}{\gamma-1} \left(T_h-T_c\right)\ln{\left(\frac{T_h}{T_c}\right)},
    \label{eq:canot}
\end{equation}
where \(\gamma\) is the specific heat capacity ratio and \(k_B\) is Boltzmann constant. For a single ($N=1$) monoatomic ideal gas (\(\gamma=\frac{5}{3}\)), the work done in the triangular cycle (\(W_\text{CT}\)) from Eqs.\eqref{eq:Workmonotri} and \eqref{eq:canot} is then given by,

\begin{equation}
    W_\text{CT} = \frac{3k_B}{4} \left(T_h-T_c\right)\ln{\left(\frac{T_h}{T_c}\right)}.
    \label{eq:workmonoatomic}
\end{equation}
At the end of the cycle using Eq.\eqref{heatclassical}, Eq.~\eqref{eq:secondall} becomes,

\small
\begin{equation}
    \begin{split}
     &\quad \Delta S_\text{sys}=0,\quad \text{and} \quad\Delta S_\text{tot}=\Delta S_\text{bath} = -\frac{Q_{\text{hot}}}{T_h}-\frac{Q_\text{cold}}{T_c},\\
    & \Delta S_\text{tot} = (S_A-S_B) + (S_B-S_A)\frac{T_h+T_c}{2T_c} = (S_B-S_A)\,\frac{T_h-T_c}{2T_c}.
    \end{split}
    \label{eq:12}
\end{equation}
\normalsize

\noindent The second law is not violated if \(Q_\text{hot}\) and \(Q_\text{cold}\) obey the condition Eq.~\eqref{eq:secondlawcond},  

\begin{equation}
    \frac{Q_{\text{hot}}}{T_h}+\frac{Q_\text{cold}}{T_c}= (S_A-S_B)\,\frac{T_h-T_c}{2T_c} \;\leq\; 0 .
\end{equation}

Thus, the classical triangular does not violate the second law if \(S_A \le S_B\), which gives us \(W \ge 0\), \(Q_{\text{hot}} \ge0 \), and \(Q_\text{cold}\le0\), therefore the triangular cycle can only operate as a heat engine. The efficiency of the classical triangular cycle as a heat engine is then given as,

\begin{equation}
    \eta_\text{CT} = \frac{W_\text{CT}}{Q_{\text{hot}}} = \frac{T_h - T_c}{2T_h} = \frac{\eta_c}{2}
    \label{eq:trianeta}
\end{equation}

\noindent where \(\eta_c = 1- \frac{T_c}{T_h}\), is the Carnot efficiency. From Eqs. \eqref{eq:Workmonotri} and \eqref{eq:trianeta} we notice that the work done and efficiency of the classical triangular cycle is half that of Carnot cycle.


\subsection{Three-Stroke Classical Isochoric Cycle}
\label{IIB}

\begin{figure}
    \centering
    \includegraphics[width=0.95\linewidth]{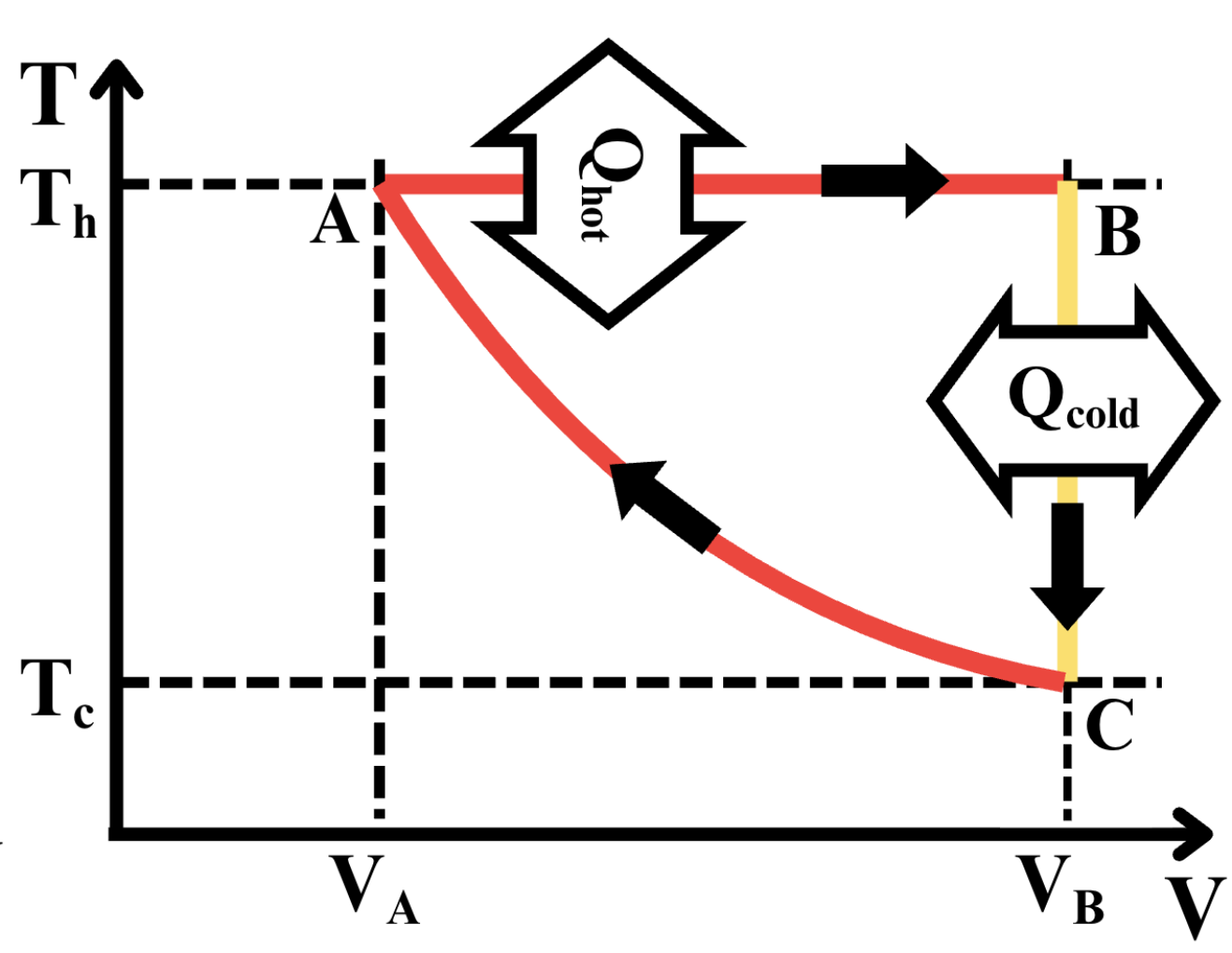}
    \caption{Volume - Temperature diagram of ideal gas during a three-stroke classical isochoric cycle. Stroke \(A\xrightarrow{}B\) is the isothermal stroke, \(B\xrightarrow{\text{Yellow}}C\) is the isochoric stroke, and \(C\xrightarrow{}A\) is the adiabatic stroke.}
    \label{Fig1.5}
\end{figure}

Fig.~\ref{Fig1.5} depicts an ideal gas undergoing a three-stroke classical isochoric cycle. Stroke \(A \rightarrow B\) is isothermal, where the system at temperature \(T_h\) and volume \(V_A\) expands to volume \(V_B\) while maintaining equilibrium with the hot bath at temperature \(T_h\). During this process, the system exchanges heat \(Q_{\text{hot}}\) with the hot bath. Stroke \(B \xrightarrow{\text{Yellow}} C\) is the classical isochoric process, where the volume is kept fixed at \(V_B\) as the system cools to temperature \(T_c\) of the cold bath. No work is done during this stroke, and the change in internal energy is entirely due to heat exchange \(Q_{\text{cold}}\) with the cold bath. Finally, in stroke \(C \rightarrow A\), the system returns to its initial state through an adiabatic process, where no heat is exchanged and only work is performed. Using standard thermodynamic relations, we compute the heat exchanged with the hot and cold baths. For a classical isothermal process, the internal energy remains constant. Thus, from Eq.~\eqref{eq:firstlaw}, we obtain \cite{callen2006thermodynamics,landau2013statistical},
\begin{equation}
    Q_{\text{hot}} = W_{\text{Isothermal}} = N k_B T_h \ln{\frac{V_B}{V_A}},
    \label{eq:classic isochor qht}
\end{equation}
where \(N\) is the number of atoms and \(k_B\) is the Boltzmann constant. For a classical isochoric process, no work is exchanged. Therefore, from Eq.~\eqref{eq:firstlaw}, we obtain \cite{callen2006thermodynamics},
\begin{equation}
    Q_{\text{cold}} = \Delta U = C_v (T_c - T_h) = \frac{N k_B (T_c - T_h)}{\gamma - 1} ,
    \label{eq:classic isochor qcld}
\end{equation}
where \(C_v\) is the molar specific heat capacity at constant volume and \(\gamma\) is the specific heat capacity ratio. The adiabatic stroke \(C \rightarrow A\) yields the relation \cite{callen2006thermodynamics,landau2013statistical},
\begin{equation}
    T_c V_B^{\gamma - 1} = T_h V_A^{\gamma - 1} \Rightarrow \frac{V_B}{V_A} = \left( \frac{T_h}{T_c} \right)^{\frac{1}{\gamma - 1}}.
    \label{eq:classic isochor adia}
\end{equation}

To verify the feasibility of the cycle, we substitute Eqs.~\eqref{eq:classic isochor qht} and \eqref{eq:classic isochor qcld} into Eq.~\eqref{eq:secondlawcond},
\begin{equation}
\begin{split}
    \frac{Q_{\text{hot}}}{T_h} &+ \frac{Q_{\text{cold}}}{T_c} = \frac{N k_B T_h \ln{\frac{T_h}{T_c}}}{T_h (\gamma - 1)} + \frac{N k_B (T_c - T_h)}{T_c (\gamma - 1)} \\
    &= \frac{N k_B}{\gamma - 1} \left( \ln{\frac{T_h}{T_c}} - \frac{T_h}{T_c} + 1 \right) \le 0.
\end{split}
\label{classical isochor work}
\end{equation}

For a stable thermodynamic system, \(\gamma > 1\), and the quantity \(\left( \ln{\frac{T_h}{T_c}} - \frac{T_h}{T_c} + 1 \right)\) is always negative when \(T_h > T_c\). Hence, the cycle is consistent with the second law. For one complete cycle,
\begin{equation}
    \begin{split}
        &\Delta S_{\text{sys}} = 0, \quad \text{and} \quad \Delta S_{\text{tot}} = \Delta S_{\text{bath}} = -\frac{Q_{\text{hot}}}{T_h} - \frac{Q_{\text{cold}}}{T_c}, \\
        &\Delta S_{\text{tot}} = \frac{N k_B}{\gamma - 1} \left( \frac{T_h}{T_c} - \ln{\frac{T_h}{T_c}} - 1 \right) \ge 0.
    \end{split}
\end{equation}

Using Eqs.~\eqref{eq:classic isochor qht}, \eqref{eq:classic isochor qcld}, and \eqref{eq:classic isochor adia}, the total work exchanged during the complete classical isochoric cycle (\(W_{\text{CI}}\)) for a single (\(N = 1\)) monoatomic (\(\gamma = \frac{5}{3}\)) ideal gas is,
\begin{equation}
\begin{split}
    W_{\text{CI}} &= Q_{\text{hot}} + Q_{\text{cold}} = \frac{N k_B \left( T_h \ln{\frac{T_h}{T_c}} + (T_c - T_h) \right)}{\gamma - 1} \\
    &= \frac{3}{2} k_B \left( T_h \ln{\frac{T_h}{T_c}} + (T_c - T_h) \right).
\end{split}
\end{equation}

The efficiency (\(\eta_{\text{CI}}\)) of the classical three-stroke isochoric cycle operating as a heat engine is then given as,
\begin{equation}
\begin{split}
    \eta_{\text{CI}} = \frac{W_{\text{CI}}}{Q_{\text{hot}}} &= 1 + \frac{ \frac{N k_B (T_c - T_h)}{\gamma - 1} }{ \frac{N k_B T_h \ln{\frac{T_h}{T_c}}}{\gamma - 1} }, \\
    &= 1 + \frac{T_c - T_h}{T_h \ln{\frac{T_h}{T_c}}} = 1 + \frac{\eta_c}{\ln(1 - \eta_c)},
\end{split}
\label{eq:classical isocho eta}
\end{equation}
where \(\eta_c = 1 - \frac{T_c}{T_h}\) is the Carnot efficiency. The efficiency of the classical isochoric cycle lies between those of the Carnot and classical triangular cycles, i.e., \(\eta_\text{CT} \le \eta_{\text{CI}} \le \eta_c\). In the following section, we introduce quantum three-stroke cycles where the second stroke is replaced by either a quantum isoenergetic or quantum isochoric process. The efficiency and work output for the classical triangular and classical isochoric cycles are compared with the quantum isochoric and quantum isoenergetic cycles in Figs.~\ref{Fig3} and \ref{Fig5} in Sections~\ref{IIB} and \ref{IIC}.

\subsection{Three-Stroke Quantum Isochoric Cycle for Quantum Particle in an Infinite Potential Well}
\label{IIC}

\begin{figure}[]
    \centering
    \includegraphics[width=0.95\linewidth]{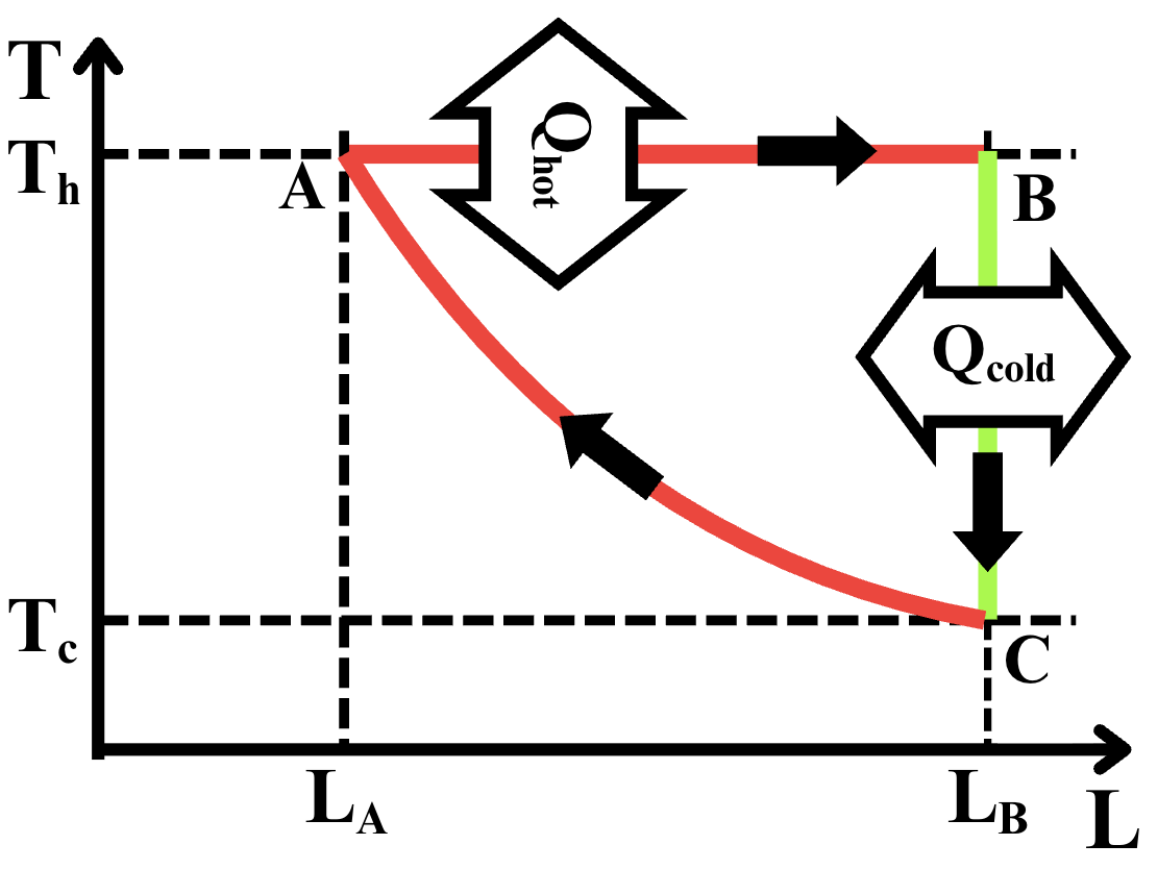}
    \caption{Length - Temperature diagram of the particle in an IPW during a three-stroke quantum isochoric cycle. Stroke \(A\xrightarrow{}B\) is the isothermal stroke, \(B\xrightarrow{\text{Green}}C\) is the quantum isochoric stroke, and \(C\xrightarrow{}A\) is the adiabatic stroke.}
    \label{Fig2}
\end{figure}

We consider a particle in a 1D IPW of variable length \(L\), operating as a three-stroke quantum isochoric thermodynamic cycle, as depicted in Figs.~\ref{Fig1} and \ref{Fig2} \((A \to B \xrightarrow{\text{Green}} C \to A)\). This cycle operates between two thermal baths at temperatures \(T_h\) and \(T_c\) (\(T_h > T_c\)), and consists of an isothermal stroke \((A \to B)\), followed by a quantum isochoric stroke \((B \xrightarrow{\text{Green}} C)\), and finally an adiabatic stroke \((C \to A)\). The quantized energy levels of a quantum particle in the 1D IPW are given by \cite{landau2013quantum,dirac1981principles},

\begin{equation}
    E_n(L)=\frac{n^2\hbar^2\pi^2}{2mL^2},
\end{equation}

\noindent where \(m\) is the particle mass (here taken to be the electron mass). For a system in thermal equilibrium at temperature \(T\), with energy eigenvalues \(E_n\) and corresponding eigenstates \(\ket{n}\), the thermal state (\(\rho(L,T)\)) is given by \cite{vinjanampathy2016quantum,quan2007quantum,quan2009quantum},
\begin{equation}
\begin{split}
        \rho(L,T) = \sum_n& p_n \ket{n}\bra{n},\quad \text{with} \quad p_n(L,T) = \frac{e^{-\beta E_n(L)}}{Z(L,T)},\\ & \text{and} \quad Z(L,T) = \sum_n e^{-\beta E_n(L)},
    \label{eq:thermal}
\end{split}
\end{equation}

\noindent where \(p_n(L,T)\) are the occupation probabilities, \(Z(L,T)\) is the partition function and \(\beta = \frac{1}{k_B T}\), where \(k_B\) is the Boltzmann constant. In this framework, temperature \(T\) is defined as the statistical parameter that establishes the system's equilibrium energy distribution through the Boltzmann distribution  \cite{quan2007quantum,pena2020otto}. For all thermodynamic cycles considered in this paper, this definition is consistently applied. During the isothermal stroke, the system remains in a Boltzmann state at \(T_h\). It then relaxes to a Boltzmann state at \(T_c\) during the isochoric stroke. Finally, the adiabatic stroke returns the system quasi-statically to its initial Boltzmann state at \(T_h\). Throughout all strokes, the system remains in a thermal state, ensuring that the temperature is well-defined at every stage. Once the thermal state is known, we can calculate other thermodynamic quantities such as entropy \(S = -k_B \sum_n p_n \ln(p_n)\) and internal energy \(U = \sum_n p_n E_n\). 

The quantum nature here arises solely from the discrete energy levels of the system, with the system described by the thermal state of Eq.\eqref{eq:thermal}, at all points in the cycle. Since this state possesses no off-diagonal elements in the energy eigenbasis, quantum coherence is absent by design, as also studied in Refs.~\cite{quan2007quantum,dann2020quantum,pena2020otto}. In this idealized, quasi-static limit, coherence does not play a role, the quantum adiabatic stroke is performed slowly enough to avoid transitions, preserving the diagonal, thermal structure of the state, and the heat exchange strokes maintain perfect, instantaneous equilibrium with their respective baths. The thermodynamic performance, including efficiency and work output, are governed solely by the discrete energy levels and the populations of these discrete energy levels.

For infinitesimal transformations, from Eq.~\eqref{eq:firstlaw}, we obtain,

\begin{equation}
    Q = \sum_n E_n(L) \, dp_n, \quad W = -\sum_n p_n(L,T) \, dE_n
    \label{eq:QandW}
\end{equation}

\begin{itemize}
    \item \textbf{Stroke A $\rightarrow$ B (Isothermal):} The system is coupled to the hot bath at temperature $T_h$. The well length expands from $L_A$ to $L_B$ at constant temperature. Heat exchanged during this process \(Q_{\text{hot}}\) is given by Eq.\eqref{eq:QandW},
    \begin{equation}
     Q_{\text{hot}} = \sum_n \int_{L_A}^{L_B} E_n(L) \frac{\partial p_n(L,T)}{\partial L} \, dL,
    \end{equation}
    \noindent which reduces to, (see, Appendix A),
    \begin{equation}
        Q_{\text{hot}} = T_h\left[S_B(L_B,T_h)-S_A(L_A,T_h)\right].
        \label{eq:qhotisocho}
    \end{equation}




    \item \textbf{Stroke B $\xrightarrow{\text{Green}}$ C (Quantum isochoric):} The well length is fixed at $L_B$ as the system thermalizes with the cold bath at temperature $T_c$. This gives us the condition,
    \begin{equation}
        E_n^B(L_B)=E_n^C(L_B).
        \label{eq:isochorpartcodn}
    \end{equation}
    
    No work is performed, and heat \(Q_{\text{cold}}\) can be found using Eq.\eqref{eq:QandW}, as,
    \begin{equation}
        Q_{\text{cold}} =  \sum_n E_n^B(L_B)\left[p_n^C(L_B,T_c)-p_n^B(L_B,T_h)\right].
        \label{eq:qcoldisoch}
    \end{equation}

    \item \textbf{Stroke C $\rightarrow$ A (Adiabatic):} The system is brought back to its initial state adiabatically. This enforces the entropy-matching condition that determines the initial length $L_A$,
    \begin{equation}
        S_C(L_B,T_c)=S_A(L_A,T_h),
        \label{eq:partboxadia}
    \end{equation}
    \noindent since the energy levels of particle in an IPW obey the energy scaling condition, (i.e \(\frac{E_n(L)}{E_m(L)}=\text{constant},\) for all \(n,m\)) \cite{pena2020otto,quan2007quantum}, we obtain a stricter condition for Eq.\eqref{eq:partboxadia},
    \begin{equation}
       p^A_n(L_A,T_h)=p^C_n(L_B,T_c), \quad \forall n.
    \end{equation}
    Now for some arbitrary energy levels: \(n,m \ (n\ne m )\), we have,
    \begin{equation}
    \begin{split}
        &\frac{p^A_n(L_A,T_h)}{p^A_m(L_A,T_h)}=\frac{p^C_n(L_B,T_c)}{p^C_m(L_B,T_c)}\Rightarrow
        \frac{e^{-\beta_h E_n(L_A)}}{e^{-\beta_h E_m(L_A)}}=\frac{e^{-\beta_c E_n(L_B)}}{e^{-\beta_c E_m(L_B)}},\\
        &e^{-\beta_h\frac{\hbar^2\pi^2}{2mL_A^2}(n^2-m^2)}=e^{-\beta_c\frac{\hbar^2\pi^2}{2mL_B^2}(n^2-m^2)}\Rightarrow           L_A = L_B\sqrt{\frac{T_c}{T_h}}.
            \label{eq:lengthconditionforadiaisocho}
    \end{split}
    \end{equation}
\vspace{-1.5em}


\end{itemize}

For the cycle to be consistent with the second law, it must satisfy the condition in Eq.~\eqref{eq:secondlawcond},i.e,
\vspace{-1em}

\begin{align}
  \frac{Q_\text{hot}}{T_h}&+\frac{Q_{\text{cold}}}{T_c}
  = S_B-S_A
   + \sum_n \frac{E_n^B}{T_c}\bigl(p_n^C-p_n^B\bigr),\nonumber\\
  &= S_B-S_C+\sum_n \frac{E_n^B}{T_c}\bigl(p_n^C-p_n^B\bigr), 
     \quad\quad\quad\quad \text{(using Eq.\eqref{eq:partboxadia})} \nonumber,\\
  &= \sum_n p_n^B\!\left[-k_B \ln p_n^B-\frac{E_n^B}{T_c}\right]
     + \sum_n p_n^C\!\left[k_B \ln p_n^C+\frac{E_n^B}{T_c}\right], \label{eq:lastline}
\end{align}
\vspace{-2em}

\begin{flalign}
    &\text{From Eq.\eqref{eq:thermal}, we have, }\quad p_n^B=\frac{e^{-\beta_h E_n^B}}{Z_B},\quad  p_n^C=\frac{e^{-\beta_c E_n^C}}{Z_C}.&
    \label{eq:pnB,pnC}
\end{flalign}
\noindent Rearranging Eq.\eqref{eq:lastline} and using Eq.\eqref{eq:pnB,pnC} gives,

\begin{equation}
\begin{split}
\frac{Q_\text{hot}}{T_h}+\frac{Q_{\text{cold}}}{T_c}
&= \sum_n p_n^B\!\left[-k_B \ln \frac{e^{-\beta_h E_n^B}}{Z_B}-\frac{E_n^B}{T_c}\right] \\
&\quad +\sum_n p_n^C\!\left[k_B \ln \frac{e^{-\beta_c E_n^C}}{Z_C}+\frac{E_n^B}{T_c}\right], \\
&= \sum_n p_n^B\!\left[\frac{E_n^B}{T_h}-\frac{E_n^B}{T_c}+k_B\ln{\left(Z_B\right)}\right] \\
&\quad +\sum_n p_n^C\!\left[-\frac{E_n^C}{T_c}+\frac{E_n^B}{T_c}-k_B\ln{\left(Z_C\right)}\right],\nonumber
\end{split}
\end{equation}

\begin{figure}[]
    \centering
    \includegraphics[width=0.95\linewidth]{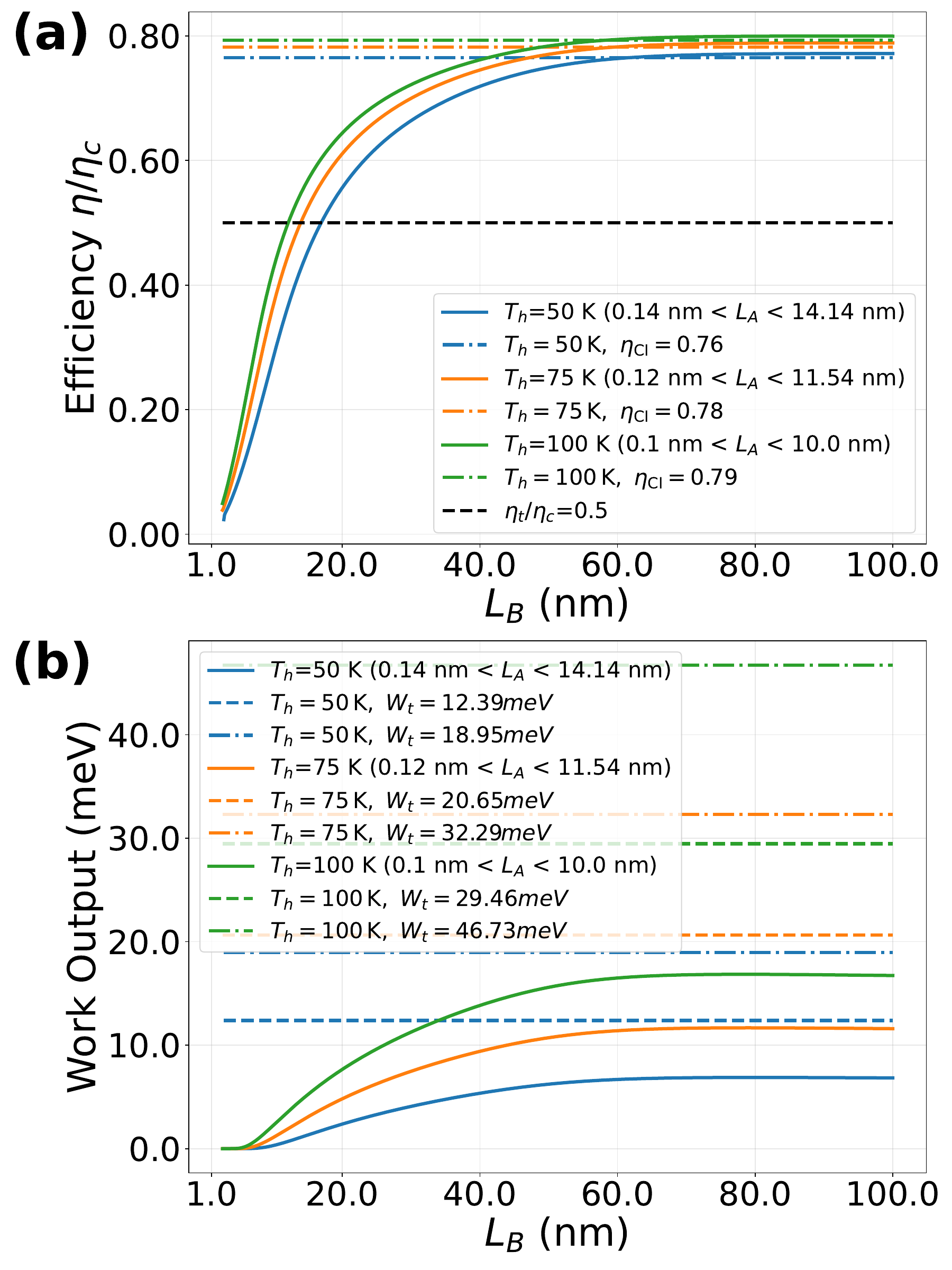}
    \caption{(a) Efficiency in units of $\eta/\eta_c$ and (b) Work output (in meV) for a particle in an IPW undergoing a three-stroke quantum isochoric cycle working as a heat engine at different hot bath temperatures. In (a), the black dashed line corresponds to $\eta_\text{CT}/\eta_c$ from Eq.~\eqref{eq:trianeta} (classical triangular cycle), while the dotted-dashed line shows $\eta_\text{CI}/\eta_c$ from Eq.~\eqref{eq:classical isocho eta} (classical isochoric cycle). In (b), the dashed and dotted-dashed lines indicate the work outputs $W_\text{CT}$ (Eq.~\eqref{eq:workmonoatomic}) and $W_\text{CI}$ (Eq.~\eqref{classical isochor work}) for a monoatomic ideal gas undergoing the classical triangular and classical isochoric cycles, respectively. $L_B$ is the length of the well during the quantum isochoric process and \(T_c=1 \text{K}\).}
    \label{Fig3}
\end{figure}

\noindent from Eq.~\eqref{eq:isochorpartcodn}, and as $\sum_np_n^B=\sum_n p_n^C=1$, the above result reduces to,

\begin{align}
\frac{Q_\text{hot}}{T_h}&+\frac{Q_{\text{cold}}}{T_c}= k_B\!\left[\ln Z_B-\ln Z_C\right]
   + \left(\frac{1}{T_h}-\frac{1}{T_c}\right)\sum_n p_n^B E_n^B\nonumber, \\
  &= -k_B\!\left[\ln\!\left(\frac{Z_C}{Z_B}\right) 
      + (\beta_c-\beta_h)\sum_n p_n^B E_n^B\right] \nonumber,\\
&=-k_B\!\left[\sum_n p_n^B\left(\ln\!\left(\frac{Z_C}{Z_B}\right) 
      +\ln(e^{(\beta_c-\beta_h)E_n^B})\right)\right]\nonumber ,\\
  &=-k_B\sum_n p_n^B\ln\!\left(\frac{e^{(\beta_c-\beta_h)E_n^B}}{Z_B/Z_C}\right) = -k_B \sum_n p_n^B\ln\!\left(\frac{\frac{e^{-\beta_hE_n^B}}{Z_B}}{\frac{e^{-\beta_cE_n^B}}{Z_C}}\right)
     \nonumber
\end{align}

\begin{equation}
= -k_B \sum_n p_n^B \ln\!\left(\frac{p_n^B}{p_n^C}\right)= -k_B \,\mathcal{D}\!\left(p^B\|p^C\right) \le 0. 
\label{eq:icho2nd}
\end{equation}

\noindent  \(p_n^i,E_n^i, S_i, Z_i,\) and \(U_i\) denote the thermal probabilities, energy levels,  entropy, partition function, and internal energy at point \(i \in \{A=(L_A,T_h), B=(L_B,T_h), C=(L_B,T_c)\}\) in parameter space, in Fig.\ref{Fig2}. Finally, \(\mathcal{D}(p\|q) = \sum_n p_n \ln\!\left(\tfrac{p_n}{q_n}\right)\) is the Kullback–Leibler relative entropy between two probability distributions \(p\) and \(q\), which is always non-negative \cite{landi2021irreversible,gray2011entropy}. Since Eq.~\eqref{eq:secondlawcond} is satisfied, the cycle is thermodynamically feasible. For one complete cycle, the entropy change for bath and system can be found using Eqs. \eqref{eq:secondall} and \eqref{eq:icho2nd},
\begin{equation}
    \begin{split}
        &\Delta S_\text{sys}=0,\quad\Delta S_\text{tot}=\Delta S_\text{bath} = k_B \,\mathcal{D}\!\left(p^B\|p^C\right)\ge0.
    \end{split}
    \label{eq:entoiscoho}
\end{equation}
For the three-stroke quantum isochoric cycle operating as a heat engine, the total work done by the system per cycle is given by \( W = Q_{\text{hot}} + Q_{\text{cold}} \), where the heat exchanged \( Q_{\text{hot}} \) and \( Q_{\text{cold}} \) are defined in Eqs.~\eqref{eq:qhotisocho} and \eqref{eq:qcoldisoch}. The work done, efficiency and coefficient of merit are then given as,

\begin{equation}
\begin{split}
    W &= T_h \bigl[S_B(L_B,T_h)-S_A(L_A,T_h)\bigr] \\
      &\quad - \sum_n E_n(L_B)\Bigl[p_n^B(L_B,T_h)-p_n^C(L_B,T_c)\Bigr],\ \  \text{and}\\[6pt]
    \eta &= \frac{W}{Q_{\text{hot}}} 
    = 1 - \frac{\sum_n E_n(L_B)\bigl[p_n^B(L_B,T_h)-p_n^C(L_B,T_c)\bigr]}
                 {T_h\bigl[S_B(L_B,T_h)-S_A(L_A,T_h)\bigr]},\\
    \text{with,}&\quad\text{Coefficient of Merit} = W \times \frac{\eta}{\eta_c}.
    \label{eq:isoceffiwork}
\end{split}
\end{equation}

Fig.~\ref{Fig3}(a) and (b) show the efficiency and work output, respectively, for a particle confined in an IPW operating as a three-stroke quantum isochoric heat engine. Both efficiency and work output increase with the temperature difference between the baths and tend to saturate for large well length $L_B$. The efficiency of the quantum isochoric cycle exceeds that of the classical triangular cycle and saturates at larger $L_B$ slightly above the efficiency of a classical three-stroke isochoric cycle. In the limit $T_h \to \infty$, Eq.~\eqref{eq:trianeta} gives $\eta_\text{CT} = \eta_c/2$ for the triangular cycle, while from Eqs.~\eqref{eq:classical isocho eta} and \eqref{eq:isoceffiwork} the efficiency of the classical and quantum isochoric cycle approaches the classical four-stroke Carnot efficiency. The work done by a particle in an IPW undergoing the quantum isochoric cycle does not exceed the work produced by a monoatomic ideal gas operating in the classical triangular and classical isochoric cycles between the same temperatures.

\subsection{Three-Stroke Quantum Isoenergetic Cycle for Quantum Particle in an Infinite Potential well}
\label{IID}

\begin{figure}[!htbp]
    \centering
    \includegraphics[width=0.95\linewidth]{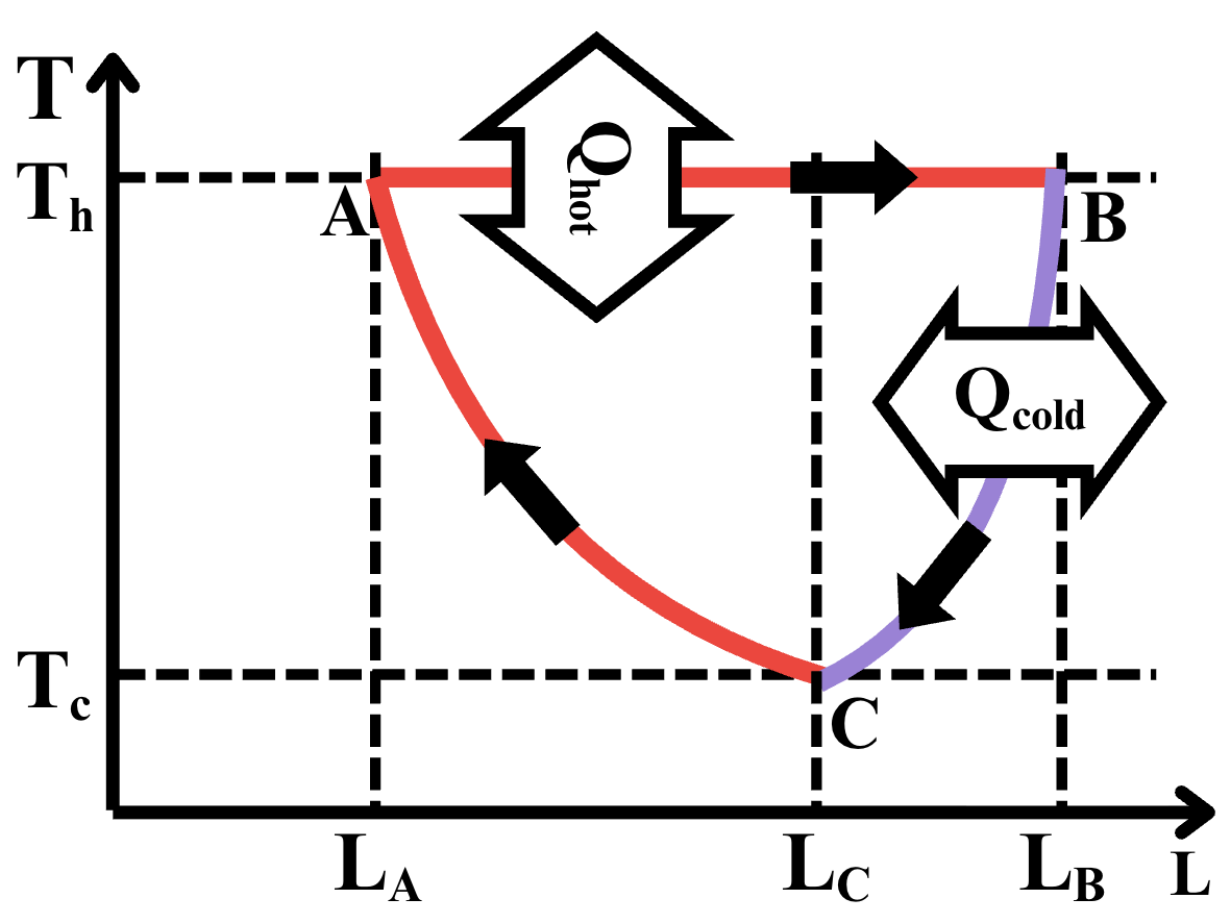}
    \caption{Length - Temperature diagram of the IPW for a three-stroke quantum isoenergetic cycle. Stroke \(A\xrightarrow{}B\) is the isothermal stroke, \(B\xrightarrow{\text{Purple}}C\) is the isoenergetic stroke, and \(C\xrightarrow{}A\) is the adiabatic stroke.}
    \label{Fig4}
\end{figure}

This cycle was first, studied in Ref.~\cite{ou2016exotic}, and considers a particle in a 1D IPW of variable length \(L\) that interacts with a single heat bath at temperature \(T_h\). The cycle (Figs.\ref{Fig1} (purple line) and \ref{Fig4}) consists of three strokes, an isothermal stroke (\(A\xrightarrow{}B\)), an isoenergetic stroke where the internal energy is held constant (\(B\xrightarrow{\text{Purple}}C\)), and an adiabatic stroke (\(C\xrightarrow{}A\)). The exact derivation for the heat exchanged and the work performed can be found in Ref.~\cite{ou2016exotic}. Here, we briefly review the cycle.  

\begin{figure}[!htbp]
    \centering
    \includegraphics[width=0.92\linewidth]{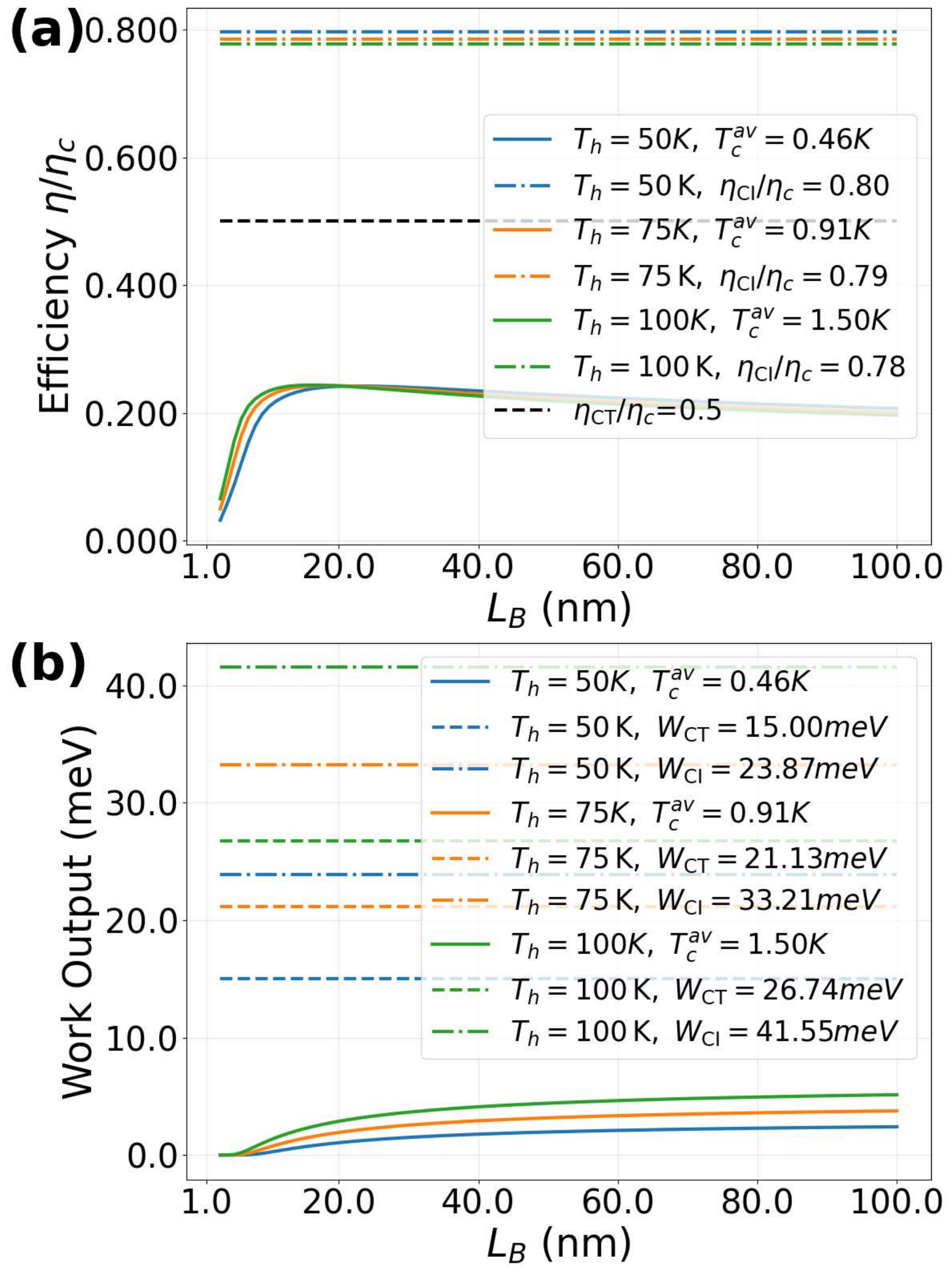}
    \caption{(a) Efficiency in units of $\eta/\eta_c$ and (b) Work output (in meV) for a particle in an IPW undergoing a three-stroke quantum isoenergetic cycle working as a heat engine at different bath temperatures. In (a), the black dashed line corresponds to $\eta_\text{CT}/\eta_c$ from Eq.~\eqref{eq:trianeta} (classical triangular cycle), while the dotted-dashed line shows $\eta_\text{CI}/\eta_c$ from Eq.~\eqref{eq:classical isocho eta} (classical isochoric cycle). In (b), the dashed and dotted-dashed lines indicate the work outputs $W_\text{CT}$ (Eq.~\eqref{eq:workmonoatomic}) and $W_\text{CI}$ (Eq.~\eqref{classical isochor work}) for a monoatomic ideal gas undergoing the classical triangular and classical isochoric cycles, respectively. Cold bath temperature used for the classical triangular and classical isochoric cycle is taken to be the average of \(T_c\) attained by the system at the end of the isoenergetic stroke within the length considered. $L_B$ is the initial length of the well during the isoenergetic process and \(L_A=1 \text{nm}\) .}
    \label{Fig5}
\end{figure}

\begin{itemize}
    \item \textbf{Stroke A $\rightarrow$ B (Isothermal):} The system, initially coupled to the hot bath at temperature $T_h$ with well length $L_A$, undergoes an expansion. The well length changes from $L_A$ to $L_B$ at constant temperature $T_h$. The heat absorbed from the hot bath is given by,
    \begin{equation}
        Q_{\text{hot}} = T_h\big[S_B(L_B,T_h)-S_A(L_A,T_h)\big].
        \label{eq:Qhot}
    \end{equation}

    \item \textbf{Stroke B $\xrightarrow{\text{Purple}}$ C (Isoenergetic):} This stroke is performed at constant internal energy, and the system is brought to an effective temperature $T_c$ while the well length changes to $L_C$. This step imposes the condition,
    \begin{equation}
        U_B(L_B,T_h)=U_C(L_C,T_c),
        \label{eq:Ucondition}
    \end{equation}
    the heat rejected (\(Q_{\text{cold}}\)) during this process is equal to the work exchanged (\(W'\)) \cite{ou2016exotic},
    \begin{equation}
        Q_{\text{cold}} = W'= \int_{L_B}^{L_C} f(L,T)dL ,
    \label{eq:Qcold}
    \end{equation}
    \noindent where \(f(L,T)=-\sum_n \frac{\partial E_n(L)}{\partial L}p_n(L,T)\), is the generalized force. For a particle in an IPW, \(\frac{\partial E_n(L)}{\partial L}=-2\frac{E_n(L)}{L}\), which implies \(f(L,T)=2\frac{U(L,T)}{L}\) (see Eq.(5) in Ref.\cite{ou2016exotic}). Since, internal energy is kept constant throughout this stroke, we obtain,
    \begin{equation}
        Q_{\text{cold}}=- 2U_B(L_B,T_h)\ln\!\left(\frac{L_B}{L_C}\right).
        \label{Qcold2}
    \end{equation}

    \item \textbf{Stroke C $\rightarrow$ A (Adiabatic):} The system returns to its initial state without heat exchange in this stroke, which obeys the adiabatic condition;
    \begin{equation}
        S_A(L_A,T_h)=S_C(L_C,T_c) ,
        \label{eq:adia_cond}
    \end{equation}
    which, following a derivation similar to Eq.\eqref{eq:lengthconditionforadiaisocho}, reduces to, (see, Eq.(12) of Ref.\cite{ou2016exotic})
    \begin{equation}
        L_C = L_A \sqrt{\frac{T_h}{T_c}}.
        \label{eq:lengthconditionforadia}
    \end{equation}
\end{itemize}

The two conditions, Eqs.~\eqref{eq:Ucondition} and \eqref{eq:lengthconditionforadia}, are solved simultaneously to determine $L_C$ and $T_c$ numerically. It should be emphasized that \(T_c\) is not the bath temperature but rather the effective temperature, the system reaches at the end of the isoenergetic process, and it depends on the parameters: $L_A,L_B$ and $T_h$. The total work performed \(W= Q_\text{hot}+Q_\text{cold}\), is obtained from Eqs.\eqref{eq:Qhot} and \eqref{Qcold2}, and the efficiency of the cycle is given by,
\begin{equation}
    \eta = \frac{W}{Q_{\text{hot}}} = 1 - \frac{2U_B(L_B,T_h)}{Q_{\text{hot}}}\ln\!\left(\frac{L_B}{L_C}\right).
    \label{eq:efficiency}
\end{equation}

 Fig.~\ref{Fig5}(a) and (b) present the corresponding efficiency and work output for a particle in an IPW under a three-stroke quantum isoenergetic cycle. In this cycle, both efficiency and work output exhibit only mild dependence on the bath temperature $T_h$, with work output showing a slight increase as $T_h$ grows. We compare the quantum isoenergetic cycle's performance with a classical triangular and classical isochoric cycle operating between $T_h$ and $T_c^\mathrm{av}$, where $T_c^\mathrm{av}$ is the average effective cold temperature for a particle in an IPW undergoing the three-stroke quantum isoenergetic cycle across the considered $L_B$ range. In this comparison, we observe that both the efficiency and the work output of the quantum isoenergetic cycle remain below those of the classical triangular and classical isochoric cycle.

\begin{table*}[]
\begin{tabular}{|c|cc|cc|}
\hline
                                                                                                                                                  & \multicolumn{2}{c|}{\textbf{\(L_B\) = 50.0 nm}}                                                                                                                                                                    & \multicolumn{2}{c|}{\textbf{\(L_B\) = 100.0 nm}}                                                                                                                                                                   \\ \cline{2-5} 
\multirow{-2}{*}{\textbf{Parameters and Performance Metric}}                                                                                      & \multicolumn{1}{c|}{\textbf{\begin{tabular}[c]{@{}c@{}}Three-Stroke\\ Quantum\\ Isochoric Cycle\end{tabular}}} & \textbf{\begin{tabular}[c]{@{}c@{}}Three-Stroke\\ Quantum\\ Isoenergetic Cycle\end{tabular}} & \multicolumn{1}{c|}{\textbf{\begin{tabular}[c]{@{}c@{}}Three-Stroke\\ Quantum\\ Isochoric Cycle\end{tabular}}} & \textbf{\begin{tabular}[c]{@{}c@{}}Three-Stroke\\ Quantum\\ Isoenergetic Cycle\end{tabular}} \\ \hline
\textbf{Length \(L_A\) in nm}                                                                                                                          & \multicolumn{1}{c|}{5.270}                                                                                     & 1.0                                                                                          & \multicolumn{1}{c|}{10.541}                                                                                    & 1.0                                                                                          \\ \hline
\textbf{\begin{tabular}[c]{@{}c@{}}Cold temperature \(T_c\) in K\end{tabular}}             & \multicolumn{1}{c|}{1.0}                                                                                       & 1.006                                                                                        & \multicolumn{1}{c|}{1.0}                                                                                       & 0.965                                                                                        \\ \hline
\textbf{\begin{tabular}[c]{@{}c@{}}Efficiency \\ (\(\eta/\eta_c\))\end{tabular}}                                         & \multicolumn{1}{c|}{{ \textbf{0.779}}}                                                     & 0.220                                                                                        & \multicolumn{1}{c|}{{ \textbf{0.796}}}                                                     & 0.197                                                                                        \\ \hline
\textbf{\begin{tabular}[c]{@{}c@{}}Work done (W)\\ in meV\end{tabular}}                                                                           & \multicolumn{1}{c|}{{ \textbf{13.610}}}                                                    & 3.905                                                                                        & \multicolumn{1}{c|}{{ \textbf{14.642}}}                                                    & 4.579                                                                                        \\ \hline
\textbf{\begin{tabular}[c]{@{}c@{}}Coefficient of Merit\\ (\(W \times \eta/\eta_c\)) in meV\end{tabular}} & \multicolumn{1}{c|}{{ \textbf{10.602}}}                                                    & 0.859                                                                                        & \multicolumn{1}{c|}{{ \textbf{11.649}}}                                                    & 0.904                                                                                        \\ \hline
\end{tabular}
\caption{Performance comparison of particle in a 1D IPW undergoing a three-stroke quantum isochoric cycle versus that of one undergoing a three-stroke quantum isoenergetic cycle for different $L_B$ values with hot bath temperature $T_h = 90\,\mathrm{K}$. The cold bath temperature for the three-stroke quantum isochoric cycle is \(T_c=1.0\ \mathrm{K}\), the length \(L_C\) for three-stroke isoenergetic cycle for the parameters $L_B=50.0\  \mathrm{nm}$ and $L_B=100.0\  \mathrm{nm}$ are $L_C =9.45\  \mathrm{nm}$ and $L_C = 9.65 \ \mathrm{nm}$ respectively. The bolded numbers highlight the three-stroke cycle  with superior performance metric for the given parameters considered.}
\label{table1}
\end{table*}

Unlike the single-bath isoenergetic cycle, where increasing $T_h$ does not enhance efficiency (see, Fig.\ref{Fig5}(a)), the three-stroke quantum isochoric cycle in Sec.\ref{IIB}, demonstrates clear gains in both efficiency and work output (see, Fig.\ref{Fig3}(a) and (b)). A performance comparison between the three-stroke quantum isochoric and isoenergetic cycles is presented in Table \ref{table1}. The temperature of the cold bath during the quantum isochoric cycle, $T_c$, is fixed at 1 K. This $T_c$ is approximately the effective cold temperature achieved in the isoenergetic cycle for the parameters considered. The results indicate that the quantum isochoric cycle achieves superior efficiency, work output and a higher overall coefficient of merit compared to isoenergetic cycle. Furthermore, implementing a quantum isochoric stroke is experimentally more feasible than realizing a quantum isoenergetic stroke~\cite{pena2016optimization,uusnakki2025experimental}. For these reasons, in the following section we concentrate on graphene-based systems undergoing the quantum isochoric cycle.

\section{Landau Levels of Graphene Systems}
\label{III}

In the presence of a perpendicular magnetic field, for MLG and BLG, the Landau level energies $E_n^{\text{MLG}}(B)$ and $E_n^{\text{BLG}}(B)$ are calculated using the low-energy effective Hamiltonians near the Dirac points ~\citep{goerbig2011electronic,python2019,mccann2013electronic,mccann2006landau}, i.e, 

\begin{equation}
\small
\begin{aligned}
  E_n^{\text{MLG}}(B) &= \pm \ v_f \sqrt{2eB\hbar n}, \quad
  E_n^{\text{BLG}}(B) = \pm \frac{\hbar eB}{m_{\text{eff}}} \sqrt{n(n - 1)}
\end{aligned}
\label{eq:spec_mono_bi}
\end{equation}

\noindent where \( m_{\text{eff}} \approx 0.035\,m_e \) the effective electron mass in BLG~\citep{mccann2006landau}. For TBG, the continuum model consists of two Dirac-like Hamiltonians for each layer, coupled via interlayer tunneling across the rotated layers~\citep{moire,python2019}. Restricting interactions to the first moiré shell yields an effective eight-band Hamiltonian~\citep{moire,bistritzer2011moire,lopes2012continuum},

\begin{equation}
\small
    \mathcal{H}^{\text{TBG}}_\theta(\mathbf{k})=\begin{bmatrix} h_{\theta/2}^{\text{MLG}}(\mathbf{k}) & T_b  & T_{tr} & T_{tl} \\ T_b^{\dagger}   & h_{-\theta/2}^{\text{MLG}}(\mathbf{k_b})&0 &0  \\ T_{tr}^{\dagger} &0 & h_{-\theta/2}^{\text{MLG}}(\mathbf{k_{tr}})&0 \\ T_{tl}^{\dagger} &0 &0& h_{-\theta/2}^{\text{MLG}}(\mathbf{k_{tl}})
    \end{bmatrix}
    \label{eq:TBG}
\end{equation}

\noindent where $\mathbf{k}$ is the momentum in the moiré Brillouin zone, and $\textbf{k}_i=\textbf{k}+\tilde{\textbf{k}}_i$ with $i \in\{\text{b,tr,tl}\}$ indexing the shifted momenta $\tilde{\textbf{k}}_i$. $h_{\theta}^{\text{MLG}}(\mathbf{k})$ denotes the monolayer Dirac Hamiltonian rotated by $\theta$~\cite{python2019,soufy2025flatband,singh2021magic}. The shifted momenta are given as,

\begin{equation}
    \tilde{\textbf{k}}_b = \kappa_{\theta} (0,-1), \quad 
    \tilde{\textbf{k}}_{tr} = \kappa_{\theta} \left( \tfrac{\sqrt{3}}{2},\tfrac{1}{2} \right), \quad 
    \tilde{\textbf{k}}_{tl} = \kappa_{\theta} \left( -\tfrac{\sqrt{3}}{2},\tfrac{1}{2} \right),
    \label{eq:Hoppingvectors}
\end{equation}

\noindent with \( \kappa_{\theta} = \frac{8\pi}{3a} \sin\left( \theta/2 \right) \), where \( a = 2.46\,\text{\AA} \) is the graphene lattice constant. The interlayer hopping matrices are given as, 

\begin{equation}
    T_b = \omega \begin{bmatrix} 1 & 1 \\ 1 & 1 \end{bmatrix}, \quad 
    T_{tr} = \omega \begin{bmatrix} e^{-i \phi} & 1 \\ e^{i \phi} & e^{-i \phi} \end{bmatrix}, \quad 
    T_{tl} = \omega \begin{bmatrix} e^{i \phi} & 1 \\ e^{-i \phi} & e^{i \phi} \end{bmatrix},
    \label{eq:Hoppingmatrix}
\end{equation}

\noindent with $\omega = 110 \ \text{meV}$ being the interlayer hopping energy and $\phi = 2\pi/3$ \cite{moire}. This Hamiltonian can be reduced to a two-band model~\citep{moire,python2019},

\vspace{-1.0em}
\begin{equation}
    \mathcal{H}_{\theta}^{\text{TBG}}(\mathbf{k \approx 0}) = \hbar v_f^* \, \boldsymbol{\sigma^* \cdot k}, \quad 
    v_f^* = v_f \frac{1 - 3 \alpha_{\theta}^2}{1 + 6 \alpha_{\theta}^2},
    \label{eq:TBG_2band}
\end{equation}
\vspace{-1.0em}

\noindent with \( v_f^* \) representing the renormalized Fermi velocity and \( \alpha_{\theta} = \omega/(\hbar v_f \kappa_{\theta}) \) defining the dimensionless interlayer coupling. The renormalized Fermi velocity \( v_f^* \) goes to zero at the magic angle \( \theta^* = 1.05^\circ \), which indicates the presence of a flat band dispersion. The complete derivation of Landau levels for TBG in a magnetic field is provided in Ref.~\cite{python2019}.

\subsection{Three-Stroke Quantum Isochoric Cycle for Graphene Based system}
\label{IIIA}

\begin{figure}[]
    \centering
    \includegraphics[width=0.95\linewidth]{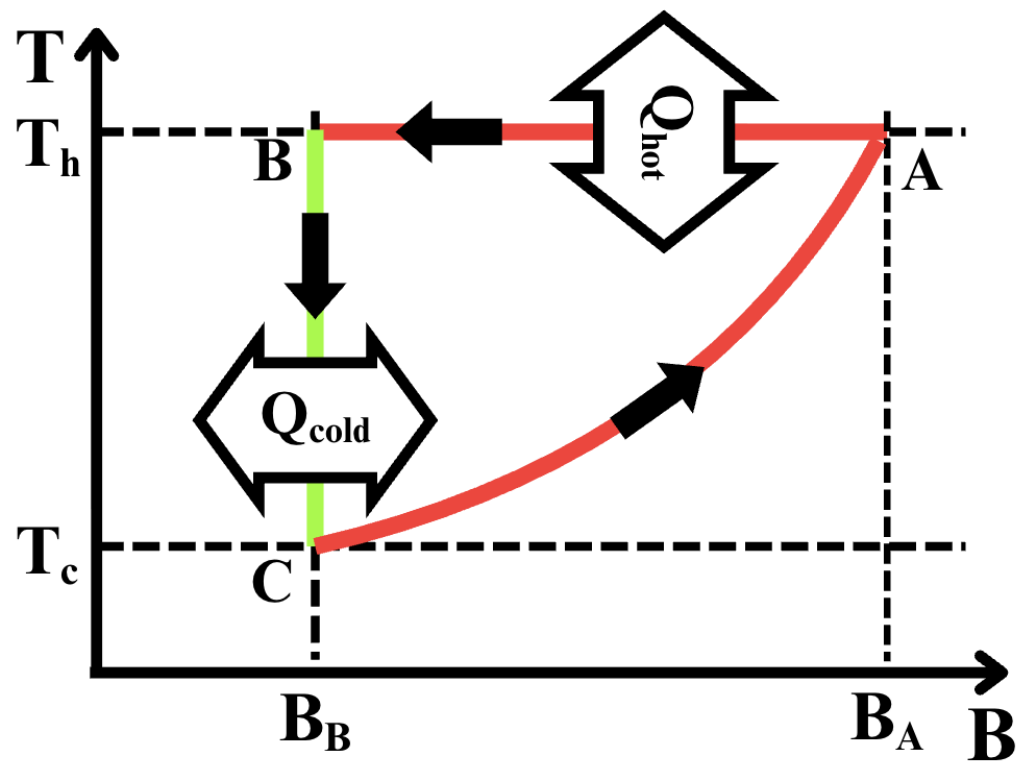}
    \caption{Magnetic field - Temperature (B-T) diagram for graphene based systems in a three-stroke quantum isochoric cycle. Stroke \(A\xrightarrow{}B\) is the isothermal stroke, \(B\xrightarrow{\text{Green}}C\) is the quantum isochoric stroke, and \(C\xrightarrow{}A\) is the adiabatic stroke.}
    \label{Fig6}
\end{figure}

Similar to a quantum particle in an IPW, a three-stroke quantum isochoric cycle can be defined for graphene based systems, with the magnetic field $B$ replacing the well length as the tunable parameter. The Landau level energies for graphene based systems can be obtained using Eqs.\eqref{eq:spec_mono_bi} and \eqref{eq:TBG}. We can then define thermal state (\(\rho(B,T)\)) for a graphene based systems as, \cite{vinjanampathy2016quantum,quan2007quantum,quan2009quantum},
\begin{equation}
\begin{split}
    \rho(B,T) = \sum_n &p_n \ket{n}\bra{n}, \quad \text{with} \quad p_n(B,T) = \frac{e^{-\beta E_n(B)}}{Z},\\ & \text{and} \quad Z(B,T) = \sum_n e^{-\beta E_n(B)},
    \label{eq:thermal_forB}
\end{split}
\end{equation}

\noindent where \(p_n(B,T)\) are the occupation probabilities, \(Z(B,T)\) is the partition function and \(\beta = \frac{1}{k_B T}\), where \(k_B\) is the Boltzmann constant. Once the thermal state is determined, other thermodynamic quantities can be computed, such as entropy  \(S = -k_B \sum_n p_n \ln p_n\) and internal energy 
\(U = \sum_n p_n E_n\). For infinitesimal transformations, using Eq.~\eqref{eq:firstlaw}, we obtain for the work done and heat exchanged,
\vspace{-0.5em}
\begin{equation}
    W = -\sum_n p_n(B,T)dE_n, \quad Q = \sum_n E_n(B)dp_n.
    \label{eq:QandW_forB}
\end{equation}

\vspace{-0.5em}

These relations are used to evaluate the performance of various graphene based systems. Now, we briefly discuss the three stroke quantum isochoric cycle for graphene based systems as depicted in Fig.\ref{Fig6}.

\begin{itemize}
    \item \textbf{Stroke A $\rightarrow$ B (Isothermal):} The system in equilibrium with the hot bath at temperature $T_h$ at magnetic field $B_A$ is brought to a magnetic field $B_B$ isothermally. The heat absorbed from the hot reservoir can be calculated using Eq.\eqref{eq:QandW_forB},
    \begin{equation}
         Q_{\text{hot}} = \sum_n \int_{B_A}^{B_B} E_n(B) \frac{\partial p_n(B,T)}{\partial B} \, dB,
    \end{equation}
    \noindent which reduces to (see Appendix A),
    
    \vspace{-2em}
    
    \begin{equation}
        Q_{\text{hot}} = T_h \left(S_B(B_B,T_h)-S_A(B_A,T_h)\right).
        \label{eq:Qhot_graphene}
    \end{equation}

    \item \textbf{Stroke B $\xrightarrow{\text{Green}}$ C (Isochoric):} The magnetic field is held constant at $B_B$ as the system thermalizes to temperature \(T_c\) of the cold bath. The heat rejected to the cold reservoir is obtained using Eq.\eqref{eq:QandW_forB},
    \begin{equation}
        Q_{\text{cold}} = \sum_n E_n(B_B)\left(p_n^C(B_B,T_c)-p_n^B(B_B,T_h)\right).
        \label{eq:Qcold_graphene}
    \end{equation}
    \vspace{-2em}
    \item \textbf{Stroke C $\rightarrow$ A (Adiabatic):} The system is decoupled from the bath and the magnetic field is changed back to its initial value $B_A$ with no heat exchange. As this is an isoentropic process, we have,
    \begin{equation}
        S_C(B_B,T_c)=S_A(B_A,T_h).
        \label{eq:adiabatic_condition_graphene}
    \end{equation}
\end{itemize}

\begin{figure}[]
    \centering
    \includegraphics[width=1\linewidth]{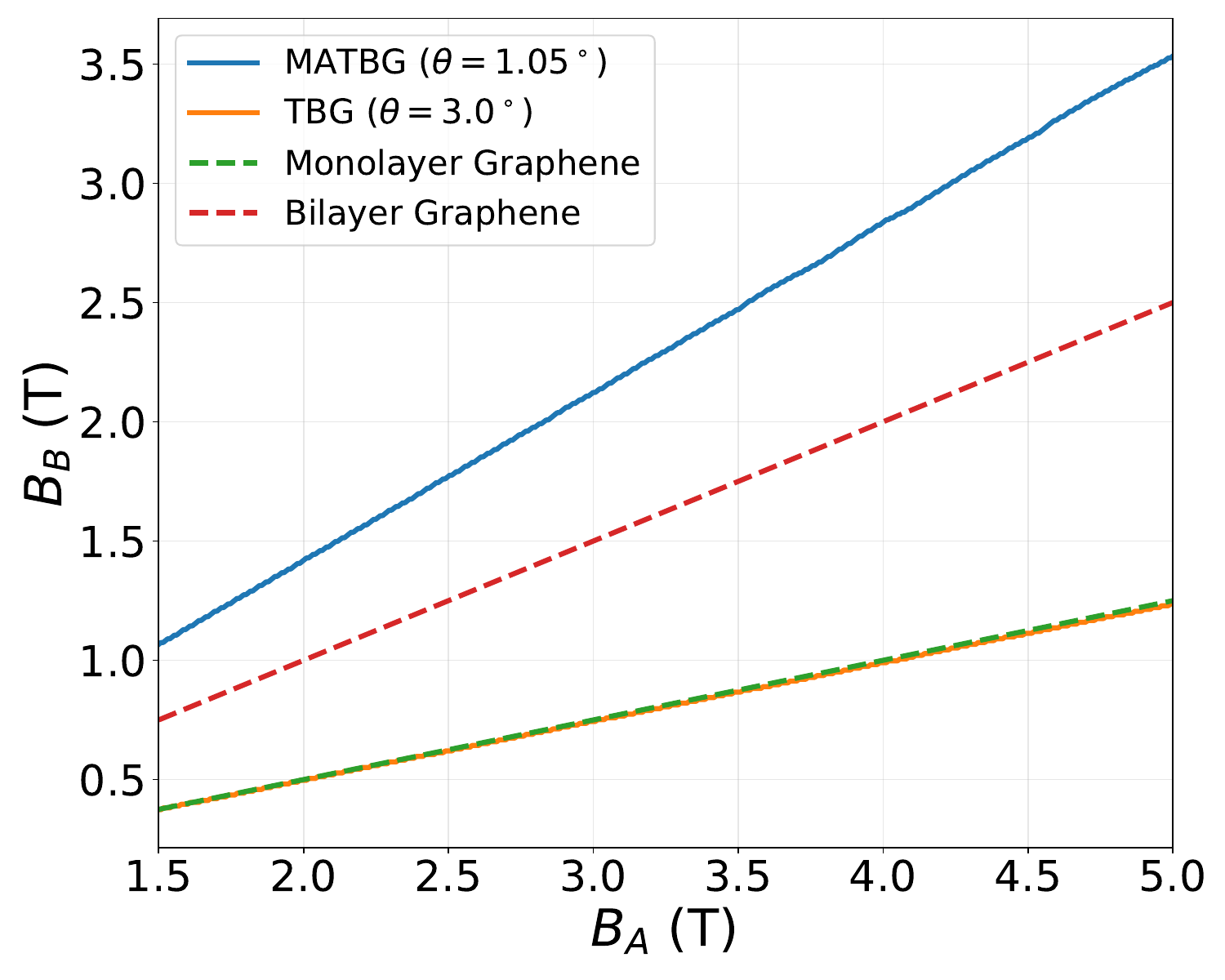}
    \caption{Adiabatic condition from Eq.~\eqref{eq:adiabatic_condition_graphene} relating the initial and final magnetic fields \(B_A\) and \(B_B\) for graphene-based systems, with bath temperatures \(T_h = 100\,\mathrm{K}\) and \(T_c = 50\,\mathrm{K}\).}
    \label{Fig7}
\end{figure}

\begin{table}[]
\begin{tabular}{|c|c|}
\hline
                                                              \textbf{Systems} & \begin{tabular}[c]{@{}c@{}}  \\ \(\alpha\)\\ \end{tabular} \\ \hline
\begin{tabular}[c]{@{}c@{}}\textbf{MLG}\\ \end{tabular}   & \(2.0000 \pm 0.0000 \approx 2.0\)                                                                                                              \\ \hline
\begin{tabular}[c]{@{}c@{}}\textbf{BLG}\\ \end{tabular}     & { \(1.0000 \pm 0.0000\approx 1.0\)}                                                                                       \\ \hline
\begin{tabular}[c]{@{}c@{}}\textbf{MATBG}\\ (\(\theta = 1.05^\circ\))\end{tabular} & \(0.4977 \pm 0.0001 \approx 0.5\)                                                                                                             \\ \hline
\begin{tabular}[c]{@{}c@{}}\textbf{TBG}\\ (\(\theta = 3.0^\circ\))\end{tabular}    & { \(2.0140 \pm 0.0002\approx 2.0\)}                                                                                       \\ \hline
\end{tabular}
\caption{Best-fit values of the exponent $\alpha$ in the relation $B_B = B_A(T_c/T_h)^\alpha$ for different graphene-based systems at $T_h = 100\,\mathrm{K}$ and $T_c = 50\,\mathrm{K}$.}
\label{table2}
\end{table}

\vspace{-1em}
As shown earlier in Sec.\ref{IIC}, Eqs.\eqref{eq:icho2nd} and \eqref{eq:entoiscoho}, the second law of thermodynamics holds for three-stroke quantum isochoric cycle. Since this derivation is general, it also applies to this cycle, yielding the same entropy changes as in Eq.~\eqref{eq:entoiscoho}. The net work done can be obtained from Eqs. \eqref{eq:Qhot_graphene} and \eqref{eq:Qcold_graphene} as \(W=Q_{\text{cold}}+Q_{\text{hot}}\). The work done and efficiency is then,  

\begin{equation}
    \begin{split}
    W&= T_h \left(S_B(B_B,T_h)-S_A(B_A,T_h)\right)\\&-\sum_n E_n(B_B)\left(p_n^B(B_B,T_h)-p_n^C(B_B,T_c)\right),\\
        \eta &=1- \frac{\sum_n E_n(B_B)\left(p_n^B(B_B,T_h)-p_n^C(B_B,T_c)\right)}{T_h \left(S_B(B_B,T_h)-S_A(B_A,T_h)\right)}.
    \end{split}
\end{equation}

For MLG and BLG, the adiabatic condition imposes stricter constraints on the thermal probabilities due to the energy-scaling condition \cite{pena2020otto,quan2007quantum,soufy2025flatband},
\begin{equation}
    p_n^C(B_B,T_c)=p_n^A(B_A,T_h) \quad \forall \ n.
    \label{eq:population_constraint}
\end{equation}

\begin{figure}[]
    \centering
    \includegraphics[width=1\linewidth]{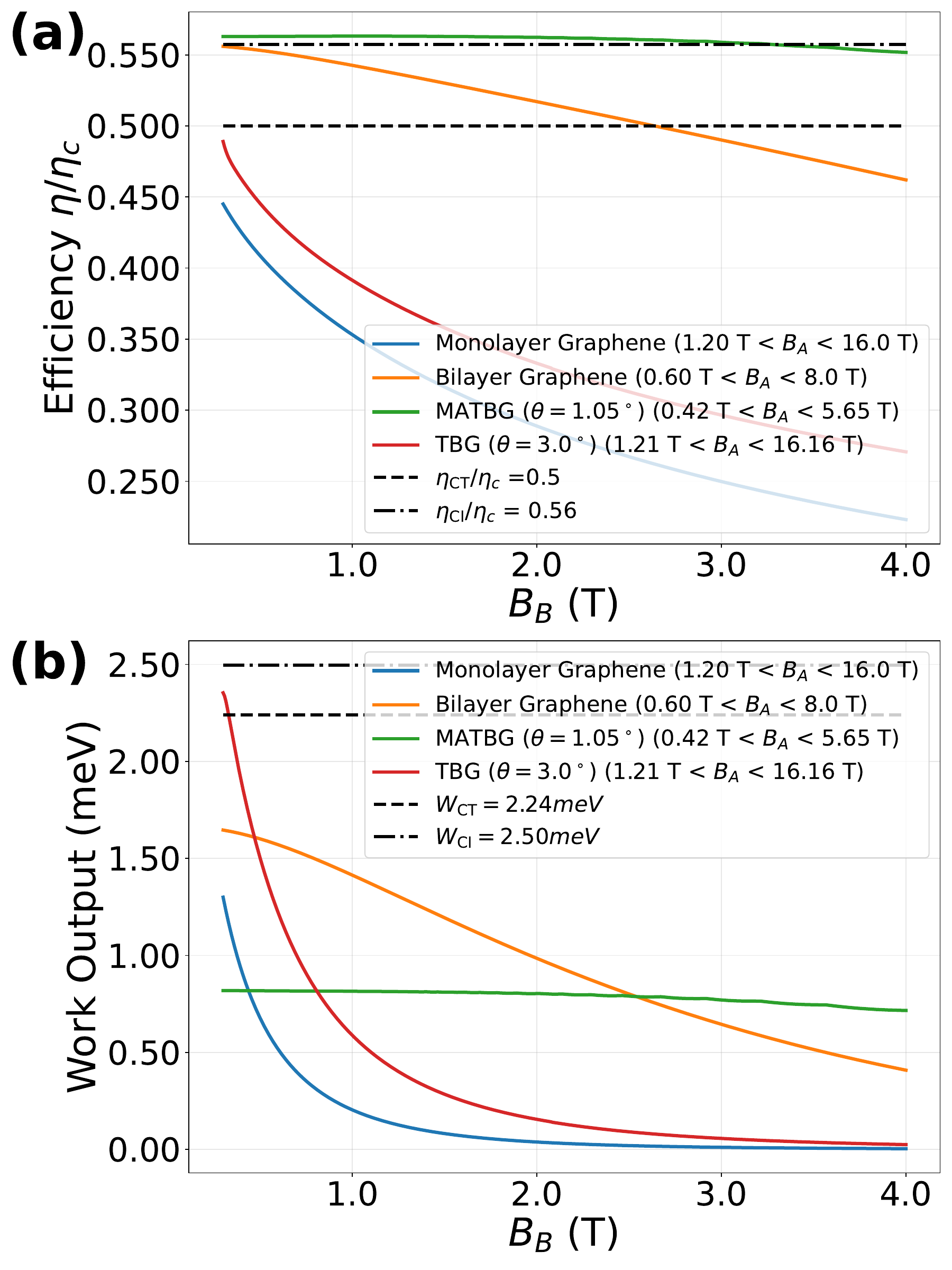}
    \caption{(a) Efficiency in units of $\eta/\eta_c$ and (b) Work output (in meV) for graphene based systems undergoing a three-stroke quantum isochoric cycle. $B_B$ is the external magnetic field during the quantum isochoric process. In (a), the black dashed line corresponds to $\eta_\text{CT}/\eta_c$ from Eq.~\eqref{eq:trianeta} (classical triangular cycle), while the dotted-dashed line shows $\eta_\text{CI}/\eta_c$ from Eq.~\eqref{eq:classical isocho eta} (classical isochoric cycle). In (b), the dashed and dotted-dashed lines indicate the work outputs $W_\text{CT}$ (Eq.~\eqref{eq:workmonoatomic}) and $W_\text{CI}$ (Eq.~\eqref{classical isochor work}) for a monoatomic ideal gas undergoing the classical triangular and classical isochoric cycles, respectively, at \(T_h=100K\) and \(T_c =50 K\).}
    \label{Fig8}
\end{figure}

Following an analogous derivation to that of Eq.~\eqref{eq:lengthconditionforadia}, we obtain explicit scaling relations linking the magnetic field to the bath temperatures,
\begin{equation}
    B_B = B_A \left(\frac{T_c}{T_h}\right)^2 \quad \text{(MLG)}, 
    \qquad 
    B_B = B_A \left(\frac{T_c}{T_h}\right) \quad \text{(BLG)}.
    \label{eq:scaling_relations}
\end{equation}

Since an analytical expression for the Landau levels of TBG cannot be obtained, we determine the relation between $B_B$ and $B_A$ using Eq.~\eqref{eq:adiabatic_condition_graphene}. Fig.~\ref{Fig7} illustrates this relation across different graphene systems. We observe that the isoentropic lines of TBG at larger twist angles (i.e, \(\theta \approx 3^\circ\)) coincide with those of MLG, reflecting the effective decoupling of TBG into MLG at large angle regime, as predicted in Refs.\cite{bistritzer2011moire,python2019}. Using a linear fit to the data in Fig.~\ref{Fig7}, we find the coefficient \(\alpha\) for the relation \(B_B = B_A(T_c/T_h)^\alpha\). Table~\ref{table2} lists the obtained values for the different graphene-based systems for \(T_h = 100\,\mathrm{K}\) and \(T_c = 50\,\mathrm{K}\).

\section{Results and discussion}
\label{IV}

In this section, we present numerical results for graphene based systems operating under a three-stroke quantum isochoric cycle. For computational accuracy, we take the first 500 Landau levels, which has been verified to yield convergent and reliable results across various graphene systems \cite{python2019,singh2021magic,soufy2025flatband}. All numerical codes used to produce the results and figures in this work are available in \cite{ThreestrokeCode}.

\begin{table}[]
\begin{tabular}{|c|cc|c|}
\hline
                                                                            & \multicolumn{2}{c|}{\textbf{Parameters}}                                                                                                                                   &                                                                                            \\ \cline{2-3}
\multirow{-2}{*}{\textbf{Systems}}                                          & \multicolumn{1}{c|}{\textbf{\begin{tabular}[c]{@{}c@{}}\(B_A\) in T\\ or \\ \(L_A\) in nm\end{tabular}}} & \textbf{\begin{tabular}[c]{@{}c@{}}\(B_B\) in T\\ or\\ \(L_B\) in nm\end{tabular}} & \multirow{-2}{*}{\textbf{\begin{tabular}[c]{@{}c@{}}Efficiency\\ (\(\eta/\eta_c\))\end{tabular}}} \\ \hline
\textbf{\begin{tabular}[c]{@{}c@{}}MLG\\ \end{tabular}}       & \multicolumn{1}{c|}{\(B_A\) = 1.756}                                                                & \(B_B\) = 0.439                                                               & 0.418                                                                                      \\ \hline
\textbf{\begin{tabular}[c]{@{}c@{}}BLG\\ \end{tabular}}         & \multicolumn{1}{c|}{\(B_A\) = 4.916}                                                                & \(B_B\) = 2.458                                                               & 0.505                                                                                      \\ \hline
\textbf{\begin{tabular}[c]{@{}c@{}}MATBG\\ (\(\theta=1.05^\circ\))\end{tabular}} & \multicolumn{1}{c|}{\(B_A\) = 1.392}                                                                & \(B_B\) = 0.986                                                               & { \textbf{0.563}}                                                      \\ \hline
\textbf{\begin{tabular}[c]{@{}c@{}}TBG\\ (\(\theta=3.0^\circ\))\end{tabular}}   & \multicolumn{1}{c|}{\(B_A\) = 3.268}                                                                & \(B_B\) = 0.809                                                               & 0.408                                                                                      \\ \hline
\textbf{\begin{tabular}[c]{@{}c@{}}Particle in \\ a 1D IPW\end{tabular}}    & \multicolumn{1}{c|}{\(L_A\) = 6.547}                                                                & \(L_B\) = 9.259                                                                & 0.536                                                                                      \\ \hline
\end{tabular}
\caption{Efficiency in units of \(\eta/\eta_c\) and parameters for graphene based systems and particle in an IPW under the three-stroke quantum isochoric cycle at identical work output  \(W = 0.815\) meV for all systems. \(\eta_c\) is the Carnot efficiency, and the bath temperatures are $T_h = 100\,\mathrm{K}$ and $T_c = 50\,\mathrm{K}$. The bolded number highlight the system with highest efficiency at equal work output, for the parameters considered.}
\label{table3}
\end{table}

Fig.~\ref{Fig8} (a) and (b) illustrate the performance of various graphene platforms operating under the three-stroke quantum isochoric cycle. We see that the thermodynamic efficiency of MATBG surpasses that of other graphene platforms, and it exceeds the classical triangular cycle efficiency across the entire range of the magnetic field and slightly surpasses the three-stroke classical isochoric cycle efficiency for lower magnetic field $B_B$ considered. Large-angle TBG and MLG display the lowest efficiency and work output, both of which decrease rapidly as $B_B$ increases. BLG on the other hand is seen to operate with modest efficiency and work output, with the efficiency dropping below the classical triangular cycle efficiency for larger values of $B_B$. From Fig.~\ref{Fig7} and Table~\ref{table2}, it is evident that the magnetic field values $B_B$ and $B_A$ are relatively closer for MATBG than for other graphene based systems operating between bath temperatures $T_h = 100$K and $T_c = 50$K. This results in a smaller work output for MATBG, as the area enclosed by the quantum three-stroke isochoric cycle will be smaller. Nevertheless, MATBG achieves higher efficiency, indicating a more effective conversion of input heat into work (see, Fig. \ref{Fig8}). Since the cycle sizes differ across graphene based systems, we compare their efficiencies at equivalent work output for a fair performance assessment. Table~\ref{table3} provides an analysis of efficiency and the required parameters for equal work output in graphene based systems and a particle in an IPW, operating under a three-stroke quantum isochoric cycle with bath temperatures $T_c = 50$ K and $T_h = 100$ K. This reveals that MATBG attains the highest efficiency at similar work output despite operating within a smaller cycle in the parameter space. The smaller area of the three-stroke quantum isochoric cycle of MATBG due to relatively close $B_B$ and $B_A$, leads to reduced heat absorption from the hot bath, as depicted in Eq.~\eqref{eq:Qhot_graphene} in turn due to a smaller entropy difference. However, MATBG rejects even less heat to the cold bath, as per Eq.~\eqref{eq:Qcold_graphene}, owing to smaller Landau level energies and spacing (see, Appendix B). These factors collectively enable MATBG to achieve higher efficiency with moderate work output.

\section{ Experimental Realization and Conclusion}
\label{V}

\begin{figure}[H]
    \centering
    \includegraphics[width=0.95\linewidth]{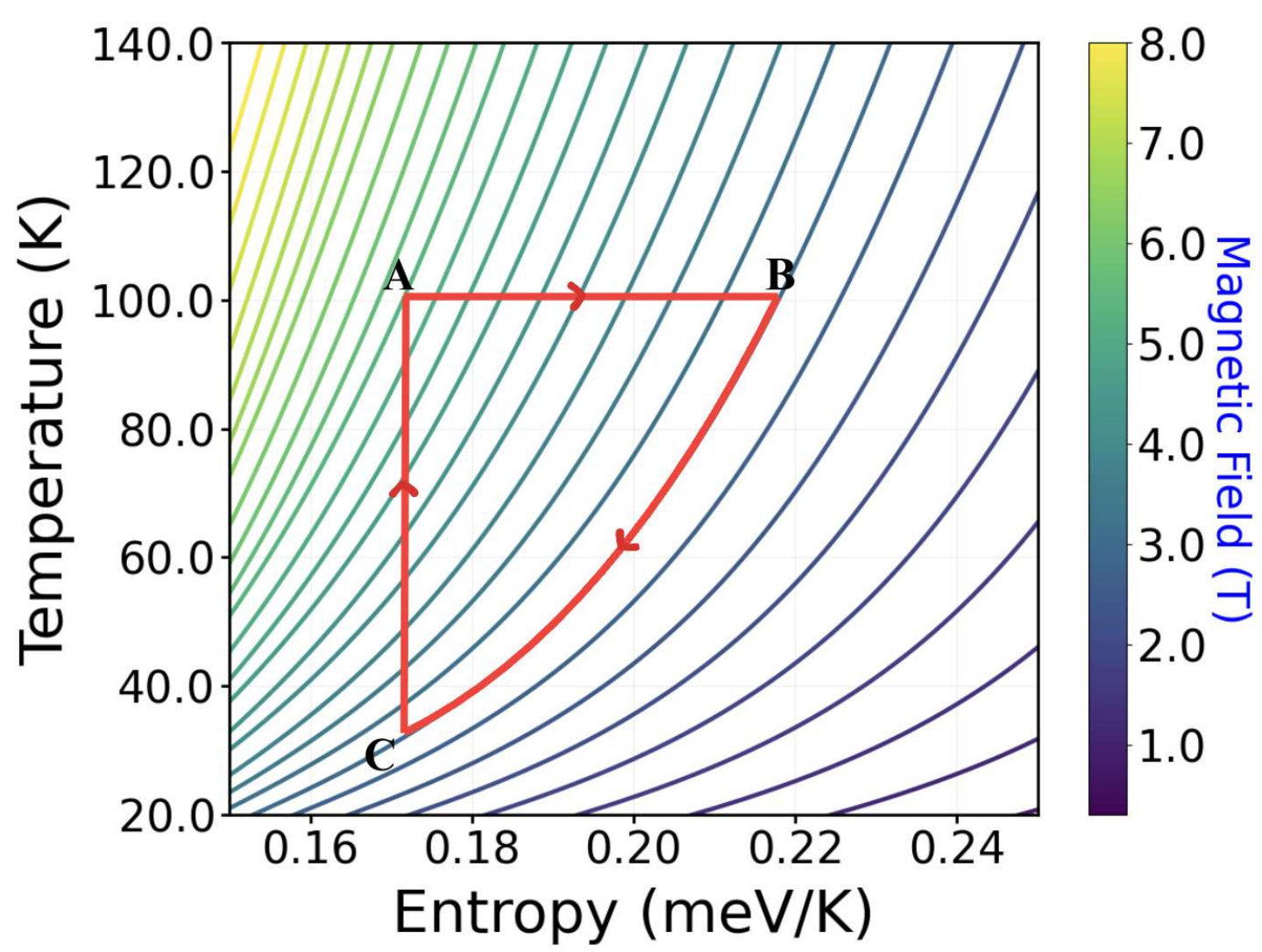}
    \caption{T-S diagram of MATBG with isochoric contour lines}
    \label{Fig9}
\end{figure}

Figure~\ref{Fig9} shows the temperature–entropy (T–S) diagram for MATBG, highlighting the isochoric contours and the three-stroke quantum isochoric cycle explored in this work. Realizing such a cycle experimentally requires precise control of both temperature and magnetic field, with feasibility determined by the ability to implement each thermodynamic stroke reliably. A quantum isochoric stroke can be achieved by coupling the system to a thermal reservoir while maintaining a constant magnetic field, an operation that is considerably more accessible experimentally than the quantum isoenergetic stroke~\cite{pena2016optimization,uusnakki2025experimental}. Few-stroke engines further reduce experimental overhead by minimizing the number of required control operations per cycle while retaining quantum-enhanced performance~\cite{klatzow2019quantum}. Recent advances in nanoscale thermodynamic measurements, such as single-electron transistor, based thermometry and scanning thermoelectric probes capable of entropy mapping~\cite{rozen2021entropic,abualnaja2023entropy,pyurbeeva2021entropy}, provide the essential tools for implementing and characterizing quantum thermodynamic cycles in MATBG and other strongly correlated two-dimensional materials.

In this work, we introduced and analyzed a three-stroke quantum isochoric cycle operating as a heat engine between two thermal reservoirs. Using a particle in a one-dimensional infinite potential well as a prototype system, we demonstrated that the cycle achieves higher efficiency than the classical triangular engine, which reaches only half the Carnot limit, and also surpasses the performance of the previously proposed three-stroke quantum isoenergetic cycle~\cite{ou2016exotic}. We then extended our analysis to graphene-based systems under perpendicular magnetic fields, including monolayer graphene (MLG), AB-stacked bilayer graphene (BLG), and twisted bilayer graphene (TBG) at both magic and non-magic angles. Among these platforms, magic-angle TBG (MATBG) achieves the highest efficiency for comparable work output, even with a smaller cycle size. This result underscores the promise of engineered quantum materials, particularly those with tunable flat-band electronic structures such as TBG, as optimal working substances for controllable, high-performance quantum thermodynamic devices.

Overall, the three-stroke quantum isochoric cycle offers distinct experimental advantages for probing and exploiting thermodynamic behavior in complex quantum materials. Its reduced operational complexity makes it well suited for nanoscale implementations, while its ability to leverage quantum effects provides a pathway toward enhanced thermodynamic performance. When combined with state-of-the-art experimental techniques, such as nanoscale thermometry, entropy imaging, and magneto-transport measurements~\cite{mecklenburg2015nanoscale}, the cycle provides a powerful framework for studying microscopic energy-conversion mechanisms in quantum systems. More broadly, our results position tunable moiré materials and low-dimensional heterostructures as versatile platforms for developing functional quantum thermodynamic technologies, including efficient nanoscale energy harvesters and integrated quantum cooling architectures.

\bibliographystyle{unsrt} 
\bibliography{ref}

\clearpage
\appendix

\subsection*{Appendix A: Heat Exchange in Quantum Isothermal Processes}
\renewcommand{\theequation}{A\arabic{equation}}  
\setcounter{equation}{0}

In quantum thermodynamics, an isothermal process is characterized by the system maintaining thermal equilibrium with a heat reservoir at a constant temperature as its external parameters change. For a quantum system subjected to a changing magnetic field \(B\), this process ensures the system evolves through instantaneous thermal states. From the first law of thermodynamics, the infinitesimal heat exchange during such a process is given by the energy-level weighted change in occupation probabilities,
\begin{equation}
    \delta Q = \sum_n E_n(B) \, dp_n(B,T),
\end{equation}
where \(E_n(B)\) are the energy eigenvalues and \(p_n(B,T)\) are the occupation probabilities. For a finite change in magnetic field from \(B_1\) to \(B_2\), the total heat exchanged becomes,
\begin{equation}
    Q = \int_{B_1}^{B_2} \sum_n E_n(B) \frac{\partial p_n(B,T)}{\partial B}  dB.
    \label{eq:appendix_heat_integral}
\end{equation}
This integral is path-dependent in parameter space, reflecting the fact that heat is a path-dependent quantity.  Now, consider the derivative of entropy ( \(S(B,T) = -k_B \sum_n p_n(B,T) \ln p_n(B,T)\)) with respect to the magnetic field,
\begin{equation}
\begin{aligned}
    \frac{\partial S}{\partial B}
    &= -k_B \sum_n \left[ \frac{\partial p_n}{\partial B} \ln p_n + p_n \frac{\partial}{\partial B}(\ln p_n) \right] \\
    &= -k_B \sum_n \frac{\partial p_n}{\partial B} \ln p_n - k_B \sum_n \frac{\partial p_n}{\partial B},
\end{aligned}
\end{equation}
where the second term vanishes due to probability conservation \(\sum_n \partial p_n/\partial B = 0\). Thus,
\begin{equation}
    \frac{\partial S}{\partial B}
    = -k_B \sum_n \frac{\partial p_n}{\partial B} \ln p_n(B,T).
    \label{eq:entropy_derivative}
\end{equation}

Using the thermal distribution expression \(\ln p_n(B,T) = -\beta E_n(B) - \ln Z(B,T)\), we substitute into Eq.~\eqref{eq:entropy_derivative},
\begin{equation}
\begin{aligned}
    \frac{\partial S}{\partial B}
    &= -k_B \sum_n \frac{\partial p_n}{\partial B} \left( -\beta E_n - \ln Z \right) \\
    &= \frac{1}{T} \sum_n \frac{\partial p_n}{\partial B} E_n + k_B \ln Z \sum_n \frac{\partial p_n}{\partial B}.
\end{aligned}
\end{equation}

The second term again vanishes due to probability conservation, yielding,
\begin{equation}
    \frac{\partial S}{\partial B}
    = \frac{1}{T} \sum_n \frac{\partial p_n}{\partial B} E_n.
    \label{eq:entropy_heat_relation}
\end{equation}

Integrating Eq.~\eqref{eq:entropy_heat_relation} from \(B_1\) to \(B_2\),
\begin{equation}
    \int_{B_1}^{B_2} \frac{\partial S}{\partial B}  dB
    = \frac{1}{T} \int_{B_1}^{B_2} \sum_n \frac{\partial p_n}{\partial B} E_n  dB.
\end{equation}

Recognizing the right-hand side as \(Q/T\) from Eq.~\eqref{eq:appendix_heat_integral}, we obtain the fundamental isothermal relation,
\begin{equation}
    S(B_2, T) - S(B_1, T) = \frac{Q}{T}
    \quad \Rightarrow \quad
    Q = T \Delta S,
    \label{eq:isothermal_heat_entropy}
\end{equation}
where \(\Delta S = S(B_2, T) - S(B_1, T)\). This result demonstrates that for quantum isothermal processes, the heat exchange is directly proportional to the entropy change, with the temperature serving as the proportionality constant. This relation holds regardless of the specific functional forms of \(E_n(B)\).

\subsection*{Appendix B: Comparative analysis of Landau level spectra and occupation probabilities in various graphene platforms}

\renewcommand{\theequation}{B\arabic{equation}}
\setcounter{equation}{0}

\begin{figure*} 
    \centering
    \begin{minipage}{0.43\textwidth}
        \centering
        \includegraphics[width=\textwidth]{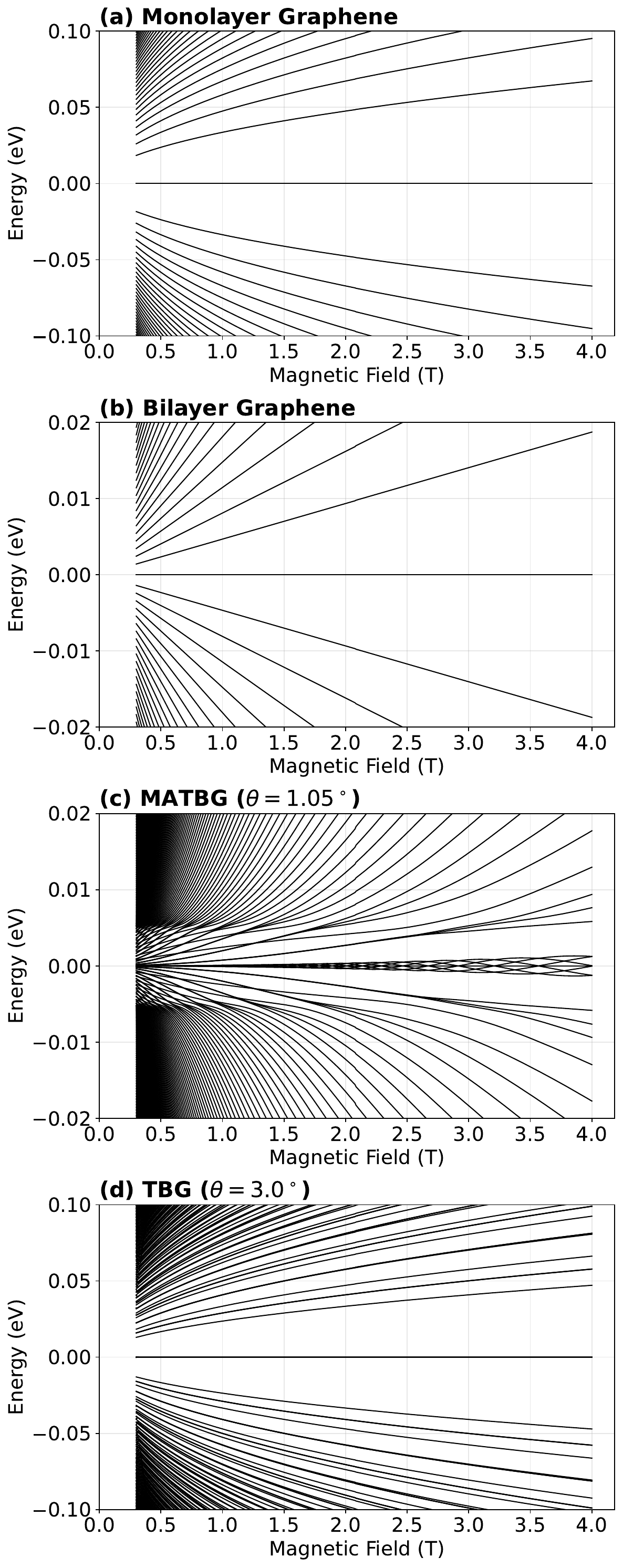}
        \caption{Magnetic field dependence of Landau level splitting in: (a) MLG, (b) BLG, (c) MATBG at $\theta=1.05^\circ$, and (d) TBG at $\theta=3.0^\circ$.}
        \label{Fig10}
    \end{minipage}
    \hspace{0.1\textwidth}
    \begin{minipage}{0.43\textwidth}
        \centering
        \includegraphics[width=\textwidth]{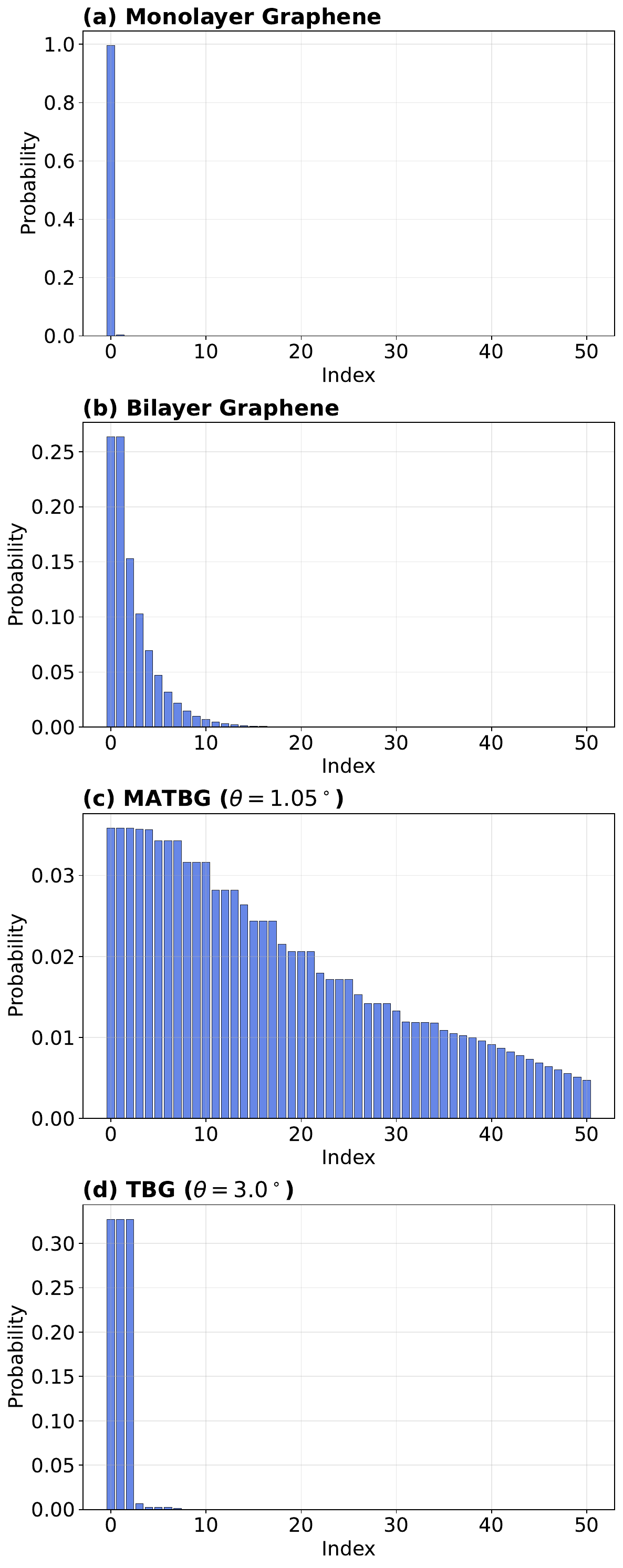}
        \caption{Occupation probabilities in various graphene platform in thermal equilibrium ($T = 50$ K, $B = 0.5$ T): (a) MLG, (b) BLG, (c) MATBG at $\theta=1.05^\circ$, and (d) TBG at $\theta=3.0^\circ$.}
        \label{Fig11}
    \end{minipage}
\end{figure*}

Fig.~\ref{Fig10} illustrates the distinct Landau level across various graphene systems, highlighting fundamental differences in their response to magnetic fields. MLG and BLG exhibit characteristic $\sqrt{B}$ and linear $B$ dependencies respectively, as seen in Eq.\eqref{eq:spec_mono_bi}, while TBG display more complex spectral features. MATBG ($\theta = 1.05^\circ$) shows exceptionally tight level packing resulting from moiré flat band formation, whereas larger twist angles ($\theta = 3.0^\circ$) progressively approach monolayer-like behavior due to interlayer decoupling \citep{python2019}. These unique Landau level spectra has profound influence over the occupation probabilities as seen in Fig.~\ref{Fig11}. MATBG's minimal level spacing and low energies enable nearly uniform occupation probability distributions, in contrast to other graphene systems where occupation is more concentrated in lower energy levels due to wider level separations.

\end{document}